\documentclass[aps,pra, twocolumn, tightenlines,amsmath,amssymb,superscriptaddress, 10pt]{revtex4-2}

\usepackage[utf8]{inputenc}
\usepackage{graphicx}
\usepackage{braket}
\usepackage{amsmath}
\usepackage{amssymb}
\usepackage{epstopdf}
\usepackage{color}
\usepackage[english]{babel}
\usepackage{lipsum}
\usepackage[dvipsnames]{xcolor}
\usepackage[normalem]{ulem}

\usepackage{siunitx}
\usepackage{booktabs}    
\usepackage[table]{xcolor} 
\usepackage[colorlinks=true, linkcolor=blue, citecolor=blue, urlcolor=blue]{hyperref}

\begin{document}


\title{Quantum Communication with Quantum Dots Beyond Telecom Wavelengths via Hollow-Core Fibers}

\author{Lorenzo Carosini}
\thanks{These authors contributed equally to this work.}
\affiliation{University of Vienna, Faculty of Physics, Vienna Center for Quantum Science and Technology (VCQ), 1090 Vienna, Austria}

\author{Francesco Giorgino}
\thanks{These authors contributed equally to this work.}
\affiliation{University of Vienna, Faculty of Physics, Vienna Center for Quantum Science and Technology (VCQ), 1090 Vienna, Austria}

\author{Patrik I. Sund}
\author{Lena M. Hansen}
\affiliation{University of Vienna, Faculty of Physics, Vienna Center for Quantum Science and Technology (VCQ), 1090 Vienna, Austria}

\author{Rene R. Hamel}
\affiliation{Optoelectronics Research Centre, University of Southampton, Southampton SO17 1BJ, UK}

\author{Lee A. Rozema}
\affiliation{University of Vienna, Faculty of Physics, Vienna Center for Quantum Science and Technology (VCQ), 1090 Vienna, Austria}

\author{Francesco Poletti}
\author{Radan Slavík}
\affiliation{Optoelectronics Research Centre, University of Southampton, Southampton SO17 1BJ, UK}

\author{Philip Walther}
\affiliation{University of Vienna, Faculty of Physics and Research Network TURIS, Boltzmanngasse 5, 1090 Vienna, Austria}
\affiliation{Institute for Quantum Optics and Quantum Information (IQOQI) Vienna,
Austrian Academy of Sciences, Boltzmanngasse 3, 1090 Vienna, Austria}

\author{Christopher Hilweg}
\affiliation{University of Vienna, Faculty of Physics and Research Network TURIS, Boltzmanngasse 5, 1090 Vienna, Austria}


\begin{abstract}
Quantum dot single-photon sources are promising for quantum communication. Yet, the most advanced devices operate near 900 nm, where standard single-mode fibers experience significant losses. We address this by employing a hollow-core fiber engineered for low-loss transmission at quantum dot wavelengths, with measured loss of 0.65 dB/km and potentially as low as 0.12 dB/km near 934 nm. The fiber also supports strong classical signals at 1550 nm without adding Raman noise. Using this platform, we transmit all four BB84 polarization states from an InAs quantum dot over 340 m with 0.1\% QBER, preserving single-photon purity and indistinguishability even in the presence of a strong classical signal. These results highlight how tailored transmission media enable quantum networks beyond the limits of standard telecom fibers.
\end{abstract}

\maketitle

\section{Introduction}

Most practical quantum communication schemes use weak coherent states or probabilistic single photons generated by spontaneous processes. Replacing these with true single photon sources would offer two significant advancements. First, the elimination of multi-photon components allows for higher key rates and removes the need for decoy-states \cite{Zhang2025,Bozzio2022}.
Second, true single-photons can in principle achieve perfect two-photon interference visibility, whereas weak coherent states are limited to \SI{50}{\percent} and heralded spontaneous parametric down conversion (SPDC) sources only reach high visibility at low pump powers, leading to an inherent trade-off between visibility and count rate. Given these difficulties, in particular in deployed situations, currently almost no quantum communication protocols make use of two-photon interference.
However, remote two-photon interference can enable qualitatively different forms of quantum communication, such as quantum relays based on single-photons \cite{zou2025realization}, quantum repeaters based on cluster states \cite{borregaard2020one}, the implementation of practical multi-partite quantum communication \cite{Proietti2021}, and provide new routes to device-independent quantum key distribution (QKD) \cite{kolodynski2020device,gonzalez2024device}.

Distributing true single-photon states through quantum networks remains challenging, as commercial standard single-mode fibers have low-loss windows of about $0.18$ dB/km and $0.32$ dB/km at \SI{1550}{\nano\meter}  and \SI{1310}{\nano\meter}, respectively. This is determined primarily by the material properties of fused silica.
These wavelengths, however, do not match those of the best single-photon sources, which typically operate at shorter wavelengths.
Among these, InAs/GaAs quantum dots (QDs) offer the highest photon count rates, lowest multi-photon emission probabilities and best two-photon interference, but they naturally emit in the \num{920}–\SI{980}{\nano\meter} range \cite{Tomm:2021aa, Ding2025, Cao2024}.
At these wavelengths, 780HP fiber is commonly used, exhibiting a loss of $~1.7$ dB/km at $930$ nm. 
Although significant progress has been achieved over the past decades \cite{Mueller2018,Yu2023,Jewon2025,Holewa2025}, \SI{1550}{\nano\meter} QDs have not yet reached the performance levels of their \SI{930}{\nano\meter} counterparts.
Consequently, recent work has focused on using nonlinear frequency conversion to shift InAs/GaAs photons to the telecom C-band \cite{zou2025realization,Morrison2021,Chiriano2025}, since transmission at their native wavelength is strongly limited by fiber losses. This approach, however, demands complex and expensive setups and inevitably introduces additional photon loss. 

Even when feasible, working with single-photons at telecom C-band wavelengths comes with its own drawbacks. Efficient detection at \SI{1550}{\nano\meter} requires costly and bulky superconducting nanowire single-photon detectors (SNSPDs), whereas Si-based detectors can achieve $\approx$\SI{75}{\percent} efficiency when operating at \SI{<1000}{\nano\meter} wavelength \cite{Acconcia2023}. Moreover, in applications requiring classical and quantum channels to coexist in the same optical fiber, (spontaneous) Raman scattering can degrade single-photon quality by transferring photons from the classical to the quantum channel~\cite{daSilva2014,Mao2018,Clivati2022}. To reduce this, the quantum channel is frequently shifted to the O-band while the classical signal remains in the C-band, lowering but not eliminating Raman noise~\cite{Grünenfelder2021}. 

\begin{figure*}[ht!]
    \centering \includegraphics[width=0.95\textwidth]{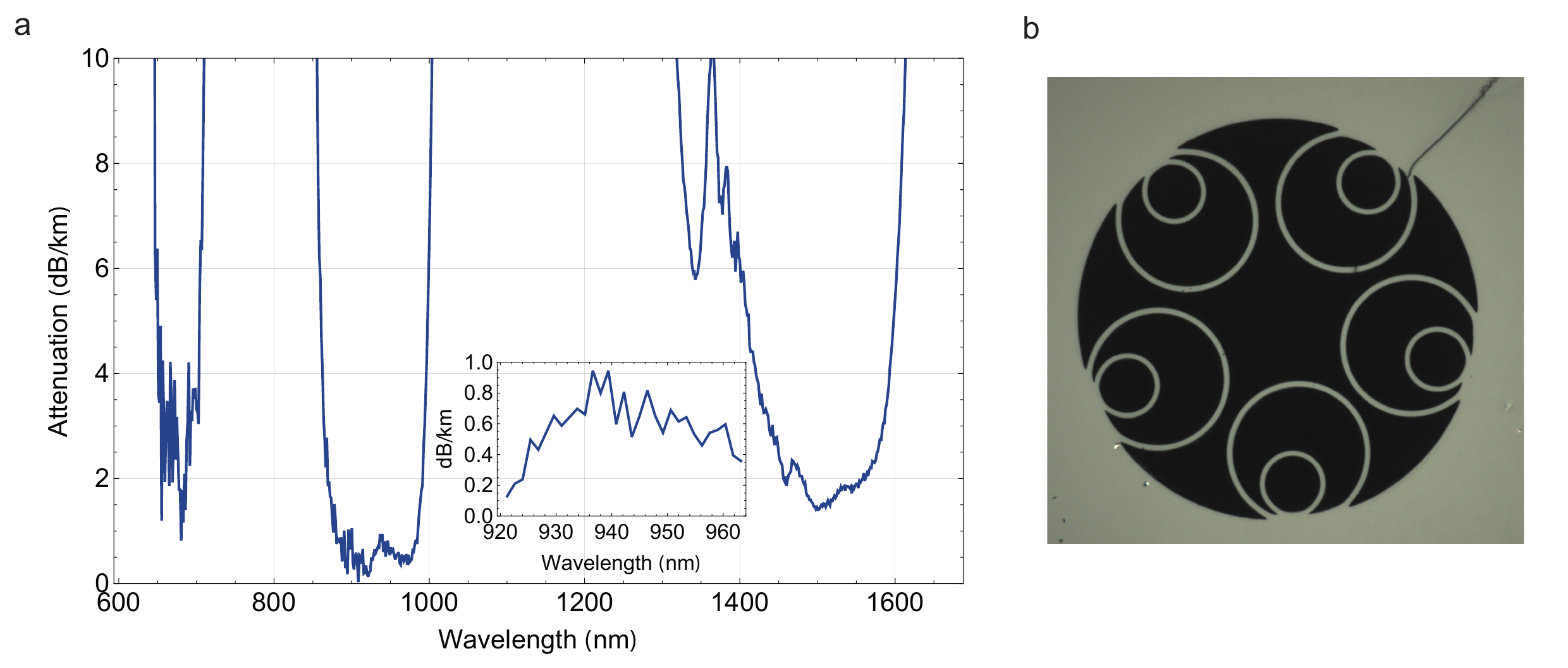}
    \caption{
\textbf{Attenuation and geometry of the AR-HCF used in this work.}
    \textbf{(a)} Wavelength-dependent attenuation of our nested anti-resonant nodeless fiber (NANF) measured using the cut-back method. The inset shows the wavelength range of interest to InAs/GaAs quantum dots.
    \textbf{(b)} Scanning Electron Micrograph (SEM). }
    \label{fig:AR-HCF}
\end{figure*}

In this letter, we propose and demonstrate a novel approach using hollow-core fibers (HCFs) with widely separated low-loss windows to transmit strong classical light at \SI{1550}{\nano\meter} alongside QD single photons at \SI{934}{\nano\meter}, with no observable degradation of the quantum channel. In HCFs, light is guided through a central hollow core rather than silica, strongly suppressing the limitations caused by light–glass interactions in standard fibers.
In fact, anti-resonant HCFs (AR-HCFs) have recently surpassed standard fibers in almost  all metrics \cite{Petrovich2025}.
Based on these exciting properties, several studies have successfully shown the coexistence of spontaneously-generated quantum light and classical light in telecom C-band~\cite{Alia2022,Honz2022,Alia2025}. Further work at telecom has used AR-HCFs to distribute spontaneously-generated entanglement~\cite{Antesberger2024,Trenti2024} and achieve superior noise performance compared to standard single mode fibers~\cite{Minder2023}.

Although AR-HCFs are typically optimized for operation over a single anti-resonant transmission wavelength window, they naturally support multiple of such windows. In our proof-of-principle experiment, we use a \SI{340}{\meter}-long AR-HCF that guides the quantum channel operating at \SI{934}{\nano\meter} in its 3rd anti-resonant window with \SI{0.65} {\decibel\per\kilo\meter} loss, and classical channels at \SI{1550}{\nano\meter} in its 2nd anti-resonant window with \SI{1.9}{\decibel\per\kilo\meter} loss (See Fig. \eqref{fig:AR-HCF}).
Although fiber design and manufacturing optimization is expected to further reduce these losses, the propagation loss in our quantum channel is already more than \SI{1}{\decibel\per\kilo\meter} lower than that achievable in standard glass-core fibers.
To show the utility of this fiber, we emulate a quantum key distribution scheme using polarization-encoded single photons from an InAs QD through the \SI{934}{\nano\meter} window while simultaneously transmitting high-power classical signals at \SI{1550}{\nano\meter} without measurable impact on the quantum channel. Our results showcase our vision of an alternative approach to quantum networks, where fibers are tailored to key quantum components but remain compatible with standard telecom channels and existing telecom components.

\begin{figure*}[ht!]
    \centering \includegraphics[width=\textwidth]{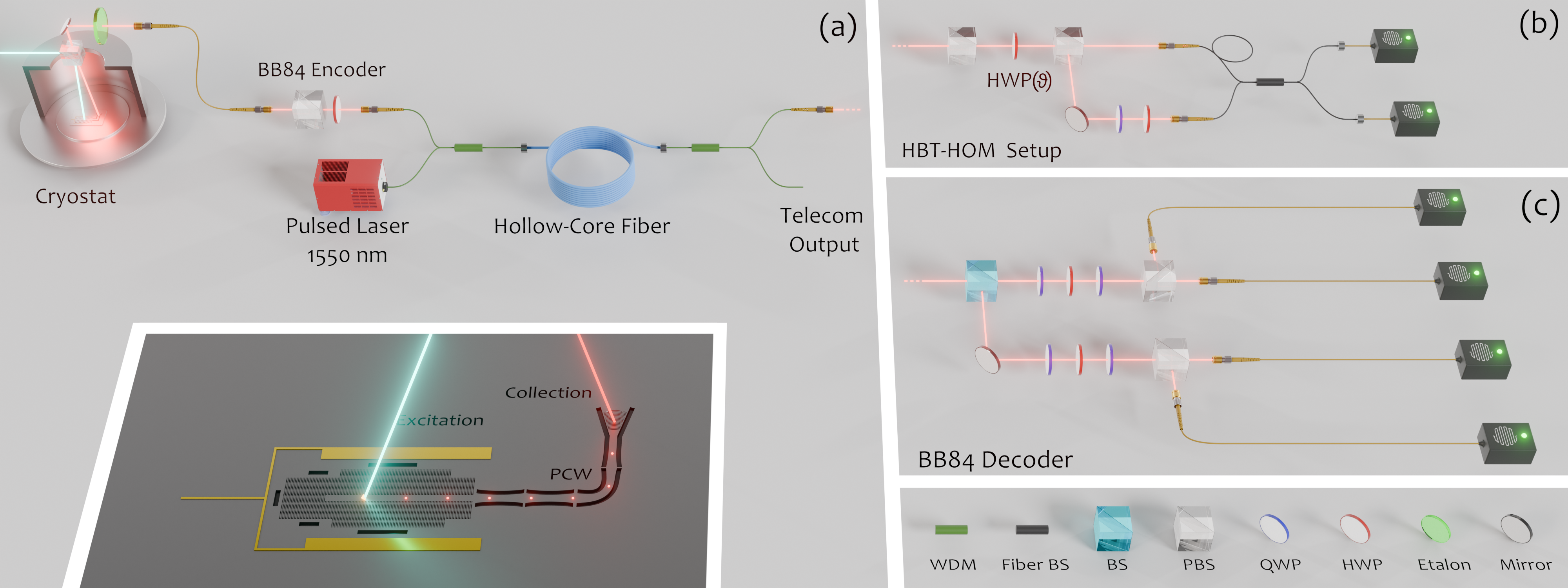}
    \caption{
    \textbf{Experimental setup for single-photon co-propagation and quantum state analysis.}
    \textbf{(a)} Emission and fiber-based transmission. The single-photon source (inset) is an InAs QD coupled to a GaAs PCW, gated via metal electrical contacts (depicted in gold), kept at 4 K inside a cryostat. Emission at 933.9~nm is filtered with a \SI{32}{\giga\hertz} bandwidth etalon to remove sidebands and combined with 53~mW of pulsed telecom light at 1550~nm via a wavelength-division multiplexer (WDM) for co-propagation through 340~m of hollow-core fiber (HCF). After the HCF, another WDM is used to separate the quantum and classical signals. The single-photon output stream is collimated in free space.
    \textbf{(b)} Single-photon purity and indistinguishability characterization. The collimated output is analyzed in a configurable setup that acts as a Hanbury Brown–Twiss (HBT) interferometer when the half-wave plate HWP($\vartheta$) is set to $\vartheta=0^\circ$, or a Hong––Ou–Mandel (HOM) interferometer at $\vartheta=22.5^\circ$. Detection is performed via superconducting nanowire single-photon detectors (SNSPDs).
    \textbf{(c)} BB84 state transmission and detection. The same output is sent to a polarization-based BB84 receiver comprising a non-polarizing beam splitter for passive basis selection, waveplates and polarizing beam splitters (PBSs) for projection, and four SNSPDs for detection. Used to evaluate quantum bit error rate (QBER) under co-propagation.
    }
    \label{figure2}
\end{figure*}

\section{Hollow-core fiber}

The light-guiding properties of AR-HCFs are primarily determined by the thin glass membranes of uniform thickness \textit{t} and refractive index \textit{n}, which act as resonators and define the wavelengths at which the fiber is effectively transparent. For nested anti-resonant nodeless fibers (NANF) the resonance wavelengths are located at ~\cite{Poletti2014}

\begin{equation}
    \lambda_m=\frac{2t}{m}\sqrt{n^2-1} \,, \hspace{1cm} m=1,2,3,...
\end{equation}

For wavelengths between these resonances, low-loss guidance can be achieved. Efficient transmission for \SI{930} and \SI{1550}{\nano\meter} over two different anti-resonant windows is possible between the resonances for $m=1/2$ (\SI{1550}{\nano\meter}) and $m=2/3$ (\SI{934} {\nano\meter}), which we used in our design. In terms of geometry, we chose a NANF design with five nested tube elements. A micrograph of the fiber manufactured using standard 2-stage stack-and-draw technique is shown in Fig.~\ref{fig:AR-HCF}b). For intended operation in the 2nd (\SI{1550}{\nano\meter}) and 3rd (\SI{934}{\nano\meter}) antiresonant windows, the tube elements thickness was designed to be \SI{1.2}{\micro\meter}. The manufactured fiber had a core diameter of \SI{27.5}{\micro\meter}, outer tubes thickness of (1.16±0.01)~$\mu m$, and nested tubes thickness of (1.20±0.02)~$\mu m$. The transmission characteristics obtained via cut-back method are shown in Fig.~\ref{fig:AR-HCF}a), where we clearly see the 2nd, 3rd, and 4th antiresonant transmission windows with minimum loss of 1.5 dBkm in the 2nd, 0.39 dB/km in the 3rd, and 2.5 dB/km in the 4th window, respectively.

In the next step, the AR-HCF was integrated into the rest of the system, which was made of components pigtailed with standard single-mode fiber. Light from the quantum (\SI{934}{\nano\meter}) and classical telecom channels (\SI{1550}{\nano\meter}) were first combined using a standard fiber-based wavelength division multiplexer (WDM), whose output was spliced to a solid core optical fiber operating at both wavelengths (SM980-5.8-125). Subsequently, it was spliced with our AR-HCF using a graded index based mode-field adapter \cite{Suslov2022}. After propagation through the AR-HCF, it was coupled back into SM980-5.8-125 using the same approach as at the input and subsequently split by an identical WDM.  

Unfortunately, the used mode-field adapter between the single-mode fiber and the AR-HCF was optimized for SMF-28 fiber and a wavelength of \SI{1550}{\nano\meter}, which is associated with significantly larger mode field diameter (\SI{10.4}{\micro\meter}) than used here (SM980-5.8-125 fiber, mode field diameter of about \SI{6}{\micro\meter} at \SI{930}{\nano\meter}). This resulted in an imperfect mode field matching with increased insertion loss, which, however, could be reduced when fabricating optimized mode field adapters (which was not available during our experiments). Together with a not fully-optimized splicing procedure and no treatment of Fresnel reflections, this produced an insertion loss of \SI{2.1}{\decibel} per SMF980-HCF interface at \SI{1550}{\nano\meter} and \SI{2.6}{\decibel} at \SI{930}{\nano\meter}. For reference, the use of an optimized procedure, including  optimized mode filed matching and an anti-reflective coating was reported to achieve loss values of \SI{0.1}{\decibel}~\cite{Zhong2024}. Thus, we expect that the loss of our quantum channel could be further reduced by as much as \SI{5.0}{\decibel} (improvement of AR-HCF-SMF980-5.8-125 connection of \SI{2.5}{\decibel} for both, input and output).

\section{Quantum signal preservation under co-propagation}

To assess the compatibility of our AR-HCF with quantum light sources, we perform transmission experiments using high-purity single photons generated by a quantum dot emitter. We will show that the AR-HCF can transmit these photons without degradation and investigate if the presence of a strong co-propagating classical signal at \SI{1550}{\nano\meter} degrades the quantum properties of the single-photon stream at \SI{930}{\nano\meter}.

\begin{figure*}
    \centering
    \includegraphics[width=\textwidth]{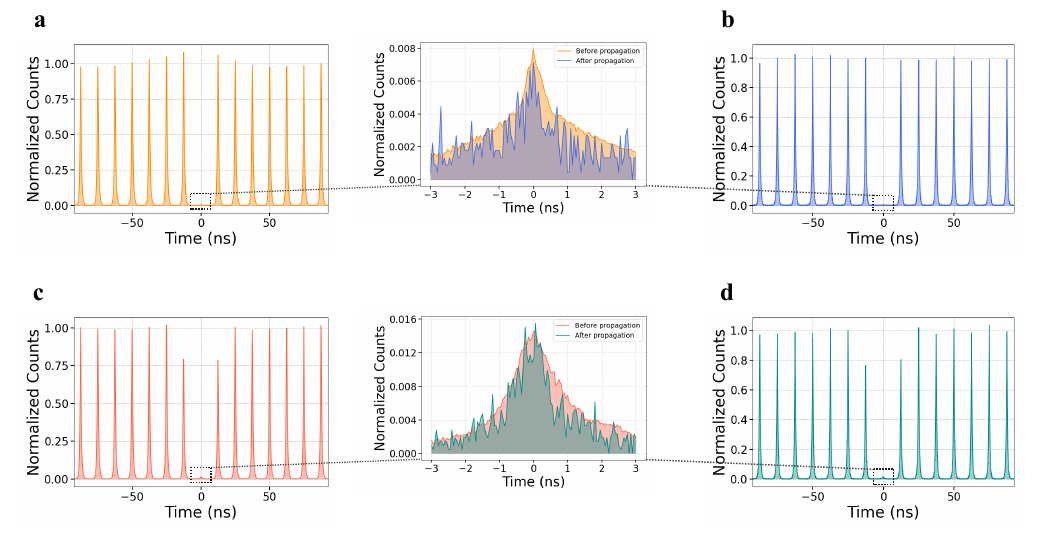}
    \caption{
    \textbf{Pre/post-fiber measurements of $g^{(2)}(0)$ and HOM visibility.}
    \textbf{a–b} Second-order autocorrelation histograms measured before (orange, left) and after (blue, right) co-propagation with a 1550~nm classical signal through 340~m of hollow-core fiber. The central inset shows normalized central peaks, highlighting minimal change. At $\pi$-pulse excitation, $g^{(2)}(0) = (21.4 \pm 0.3)\times10^{-3}$ before and $(23.3 \pm 0.7)\times10^{-3}$ after transmission.\\
    \textbf{c–d} Two-photon interference histograms from HOM measurements before (red, left) and after (cyan, right) fiber propagation. The central inset shows normalized peak suppression. HOM visibility is $(92.96 \pm 0.03)\%$ pre- and $(92.7 \pm 0.2)\%$ post-fiber. The values are computed dividing the correlated peaks area by the average of the uncorrelated areas and all areas are integrated over a 12.5~ns window without any background correction. Uncertainties have been evaluated assuming Poissonian distributed confidence levels for the involved variables.
    }
    \label{figure3}
\end{figure*}

The emitter is a single InAs quantum dot embedded in a photonic crystal waveguide (PCW), held at \SI{4}{\kelvin} in a closed-cycle cryostat (inset of Fig.~\ref{figure2}a). It is resonantly excited by a pulsed laser spectrally shaped through a folded 4-f system, to set a bandwidth of $\sim$\SI{90}{\pico\meter}, matching the QD emission at $\lambda$ = \SI{933.9}{\nano\meter}. Electrical tuning of the QD, facilitated by low-noise electrical contacts, stabilizes the charge state, ensuring emission on the desired transition and minimizing spectral diffusion due to residual charge noise~\cite{UPPU:20, pedersen2020near}. Emission is collected via a cryo-compatible objective and coupled to a single-mode fiber through a shallow-etched grating. An etalon with a bandwidth of \SI{32}{\giga\hertz} is employed as a frequency filter to optimize the indistinguishability of the emitted photons by removing the undesired phonon-induced spectral sideband. By pumping the QD with increasing pulse area Rabi oscillations are observed (see the Supplementary Text) and at $\pi$-pulse excitation we measure a single-photon rate of \SI{17.2}{\mega\hertz}, corresponding to a source fiber efficiency of \SI{25.3}{\percent} given a detection efficiency of \SI{85}{\percent}.

\begin{figure*}
    \includegraphics[width=0.9\textwidth]{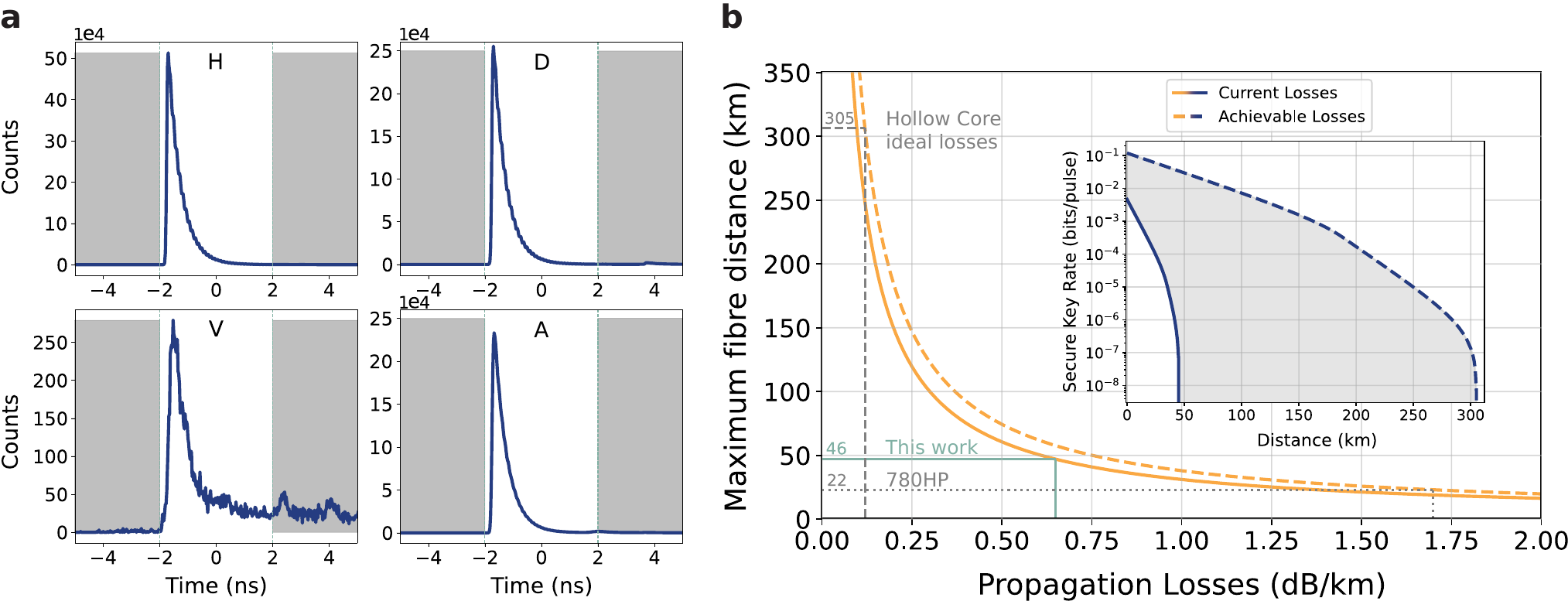}\vspace{0mm}
    \caption{\textbf{Performance of QKD test. a} Typical data set for the transmission of horizontally polarised photons showing the time gated signal to improve the QBER without compromising security. Note the scale difference between H- and V-polarized photons.
    \textbf{b} Maximum tolerable distance as a function of the propagation losses. The inset shows the asymptotic secure key rates based on experimentally measured parameters. In the plots, the solid lines have been drawn accounting for the losses measured in the system, while the performance displayed by the dashed line neglects the losses introduced by the WDM and consider ideal, yet theoretically achievable, propagation losses of \SI{0.12}{dB/\kilo\meter} and connection losses between the single mode and hollow core fibers of \SI{0.2}{dB}.} 
    \label{figure4}
\end{figure*}

A standard approach to characterize the quality of single-photon sources involves measuring their purity and indistinguishability~\cite{sps:senellart17}. These metrics respectively quantify the suppression of multiphoton events and the ability of photons to interfere, i.e. the two-photon interference visibility.
They are key benchmarks for quantum communication and quantum information processing. We perform both measurements before and after co-propagation through the hollow-core fiber to verify that transmission does not degrade the photon's quantum properties.

We perform our characterization using the modular setup shown in Fig.\ref{figure2}(b). Using a motorized half-wave plate this setup can be reconfigured between a Hanbury Brown–Twiss (HBT) interferometer, to measure the photon purity, and a Hong–Ou–Mandel (HOM) interferometer, to measure the indistinguishability.
In the HBT configuration~\cite{Brown1956}, we measure the second-order autocorrelation function, yielding $g^{(2)}(0) = 0.0214 \pm 0.0001$, confirming a strong suppression of multi-photon events. Switching to the HOM configuration~\cite{HOM87}, we evaluate two-photon interference and obtain a raw visibility of $V_{\text{HOM}} = (92.96 \pm 0.03)$\%.
These figures of merit are extracted from the coincidence histograms shown in Fig.\ref{figure3} (left panels) by integrating the areas of the central and side peaks~\cite{Loredo:16}. We highlight that the actual mean wave-packet overlap --- also referred to as \textit{indistinguishability} --- is higher because the visibility is degraded by residual multi-photon contribution \cite{Ollivier2021, GonzlezRuiz2025}. 

To test for degradation induced by co-propagation, the same quantum dot signal is injected into 340~m of HCF along with a 53~mW pulsed telecom laser at 1550~nm, using WDMs for combination and separation. The purity and indistinguishability measurements are repeated post-transmission using the same setup. As shown in Fig.\ref{figure3} (right panels), the shape and relative areas of the coincidence peaks remain largely unchanged. The extracted values show a minor increase in $g^{(2)}(0)$ to $0.0233 \pm 0.0007$ compared to the initial $0.0214 \pm 0.0003$, while the HOM visibility remains essentially unchanged within the error margins at $(92.7 \pm 0.2)\%$. These results confirm that the hollow-core fiber does not significantly degrade the non-classical nature of the emitted photons under co-propagation.

To evaluate possible leakage from the classical telecom signal into the quantum detection band, we monitor SNSPD count rates while co-propagating single photons and the telecom classical laser light. Even with up to 53~mW of classical power injected into the combining WDM, no increase in background counts is observed compared to quantum-only operation. This insensitivity arises from the low nonlinear susceptibility of the HCF and the negligible SNSPD detection efficiency at 1550~nm. The large spectral separation between the two signals enhances isolation by effectively suppressing the leakage of photons from the classical to the quantum channel.

\section{BB84 states transmission}

Having established that the quantum dot emission retains high purity and indistinguishability after transmission through the hollow-core fiber, even in the presence of strong classical signals, we proceed to assess its viability for quantum communication. To this end, we prepare and detect the four states required for the BB84 protocol for quantum key distribution, using both polarization and time-bin encodings, and evaluate the resulting quantum bit error rates (QBER) after propagation through the same fiber segment.

A schematic of the BB84 setup based on polarization is shown in Fig.~\ref{figure2}c. The single-photon stream emitted by the quantum dot is collimated into free space and passed through a PBS to prepare horizontally polarized single photons.
The four BB84 polarization states $\{\ket{H}, \ket{V}, \ket{+}, \ket{-}\}$ are prepared, multiplexed via WDMs and co-propagated through the HCF as in the previous configuration.
At the receiver end, a 50:50 beam splitter (BS) followed by free-space waveplates and polarizing BS performs passive basis analysis. The photons are coupled into four single-mode fibers and directed to four SNSPDs.
For each polarization, we execute $N=4.8 \cdot 10^9$ rounds employing a gating window of $\SI{4}{ns}$ - as shown in Fig. \ref{figure4}(a) - and measure a sifted key rate of $\approx \SI{181}{\kilo \hertz}$ with an average QBER $\approx 0.11 \%$ confirming the negligible degradation of the signal due to the fiber and the classical signal.  
In Fig.\ref{figure4}(b) we display the maximum distance for which a secure key rate (SKR) can be distilled as a function of the fiber propagation losses, together with the theoretical SKR versus channel loss, on the asymptotic GLLP bound \cite{Gottesman2004} evaluated with experimentally measured parameters. 
This shows that the fiber used in this experiment, which already outperforms the 780HP fiber commonly used at this wavelength, holds the promise to exceed even the performance of single-photon QKD at telecommunications wavelengths  \cite{Morrison2023, Zahidy2024}. 
To assess the compatibility of the system with time-bin encoding, we prepare quantum states in a BB84-like basis using a self-stabilizing interferometric scheme. Temporal qubits are encoded as superpositions of early and late time bins, with polarization used as an auxiliary degree of freedom to enable deterministic encoding and decoding. 
The average QBER measured over the four time-bin states was $\approx 0.51 \%$. This value remains well below typical security thresholds for BB84, although slightly higher than in the polarization case, due to the additional optical components involved in time-bin preparation and decoding. A detailed description of the time-bin setup, together with further details on the security analysis, is provided in the Supplementary Materials.

\section{Discussion}

This work highlights the potential of integrating hollow-core fibers with quantum dot emitters for quantum communication. Such an approach exploits the reduced multi-photon emission at high photon rates and the enhanced two-photon visibility of quantum dot photons, while avoiding the frequency conversion otherwise required for low-loss transmission in standard single-mode fibers. Moreover, the employed low-loss dual-channel AR-HCF enables simultaneous propagation of strong classical signals at telecom wavelengths without significantly degrading the quantum channel. In addition, the shorter emission wavelength of quantum dots near \SI{900}{\nano\meter} enables the use of cost-effective silicon avalanche photodiodes (APDs) instead of SNSPDs. Combined with the large spectral separation between the quantum and classical channels, this further relaxes the filtering requirements that usually constrain simultaneous O- and C-band transmission.

Looking ahead, the design flexibility of AR-HCF, and in particular NANFs and DNANFs, offers a promising path to further improvements. By optimizing the splicing to standard single-mode fibers we expect that the loss at \SI{930}{\nano\meter} can be reduced to about \SI{0.5}{\decibel} per connector, while by employing gluing approaches instead, connection losses might even reduce to \SI{0.1}{\decibel}. By engineering the fiber design parameters we can further expect propagation losses as low as \num{0.12}-\SI{0.13}{\decibel\per\kilo\meter} at \SI{930}{\nano\meter} \cite{Fokoua2023,Petrovich2025}, while simultaneously maintaining low-loss transmission at telecom wavelengths. If achieved, this would significantly improve on the secure key rates and achievable fiber distances as shown in Fig. \ref{figure4}, and provides an alternative route for future quantum networks.

\begin{acknowledgments}
This project has received funding from the European Union’s Horizon 2020 research and innovation programme under the Marie Skłodowska-Curie grant agreement No 956071 (AppQInfo) and under grant agreement no. 101017733 (QuantERA II). Co-funded by the European Union (HORIZON Europe Research and Innovation Programme, EPIQUE, No 101135288), and by the European Union (ERC, GRAVITES, No.~101071779). Views and opinions expressed are however those of the author(s) only and do not necessarily reflect those of the European Union or the European Research Council Executive Agency. Neither the European Union nor the granting authority can be held responsible for them. This work received funding from the Engineering and Physical Sciences Research Council Centre for Doctoral Training in Quantum Technology Engineering under grant EP/Y035267/1. This research was funded in whole or in part by the Austrian Science Fund (FWF) [10.55776/COE1] and [10.55776/F71] and [10.55776/FG5]. For open access purposes, the author has applied a CC BY public copyright license to any author accepted manuscript version arising from this submission. 
\end{acknowledgments}

\clearpage
\bibliography{biblio}

\begin{thebibliography}{43}%
\makeatletter
\providecommand \@ifxundefined [1]{%
 \@ifx{#1\undefined}
}%
\providecommand \@ifnum [1]{%
 \ifnum #1\expandafter \@firstoftwo
 \else \expandafter \@secondoftwo
 \fi
}%
\providecommand \@ifx [1]{%
 \ifx #1\expandafter \@firstoftwo
 \else \expandafter \@secondoftwo
 \fi
}%
\providecommand \natexlab [1]{#1}%
\providecommand \enquote  [1]{``#1''}%
\providecommand \bibnamefont  [1]{#1}%
\providecommand \bibfnamefont [1]{#1}%
\providecommand \citenamefont [1]{#1}%
\providecommand \href@noop [0]{\@secondoftwo}%
\providecommand \href [0]{\begingroup \@sanitize@url \@href}%
\providecommand \@href[1]{\@@startlink{#1}\@@href}%
\providecommand \@@href[1]{\endgroup#1\@@endlink}%
\providecommand \@sanitize@url [0]{\catcode `\\12\catcode `\$12\catcode `\&12\catcode `\#12\catcode `\^12\catcode `\_12\catcode `\%12\relax}%
\providecommand \@@startlink[1]{}%
\providecommand \@@endlink[0]{}%
\providecommand \url  [0]{\begingroup\@sanitize@url \@url }%
\providecommand \@url [1]{\endgroup\@href {#1}{\urlprefix }}%
\providecommand \urlprefix  [0]{URL }%
\providecommand \Eprint [0]{\href }%
\providecommand \doibase [0]{http://dx.doi.org/}%
\providecommand \selectlanguage [0]{\@gobble}%
\providecommand \bibinfo  [0]{\@secondoftwo}%
\providecommand \bibfield  [0]{\@secondoftwo}%
\providecommand \translation [1]{[#1]}%
\providecommand \BibitemOpen [0]{}%
\providecommand \bibitemStop [0]{}%
\providecommand \bibitemNoStop [0]{.\EOS\space}%
\providecommand \EOS [0]{\spacefactor3000\relax}%
\providecommand \BibitemShut  [1]{\csname bibitem#1\endcsname}%
\let\auto@bib@innerbib\@empty
\bibitem [{\citenamefont {Zhang}\ \emph {et~al.}(2025)\citenamefont {Zhang}, \citenamefont {Ding}, \citenamefont {Li}, \citenamefont {Zhang}, \citenamefont {Guo}, \citenamefont {Wang}, \citenamefont {Ning}, \citenamefont {Xu}, \citenamefont {Liu}, \citenamefont {Zhao}, \citenamefont {Zou}, \citenamefont {Wang}, \citenamefont {Cao}, \citenamefont {He}, \citenamefont {Peng}, \citenamefont {Huo}, \citenamefont {Liao}, \citenamefont {Lu}, \citenamefont {Xu},\ and\ \citenamefont {Pan}}]{Zhang2025}%
  \BibitemOpen
  \bibfield  {author} {\bibinfo {author} {\bibfnamefont {Y.}~\bibnamefont {Zhang}}, \bibinfo {author} {\bibfnamefont {X.}~\bibnamefont {Ding}}, \bibinfo {author} {\bibfnamefont {Y.}~\bibnamefont {Li}}, \bibinfo {author} {\bibfnamefont {L.}~\bibnamefont {Zhang}}, \bibinfo {author} {\bibfnamefont {Y.-P.}\ \bibnamefont {Guo}}, \bibinfo {author} {\bibfnamefont {G.-Q.}\ \bibnamefont {Wang}}, \bibinfo {author} {\bibfnamefont {Z.}~\bibnamefont {Ning}}, \bibinfo {author} {\bibfnamefont {M.-C.}\ \bibnamefont {Xu}}, \bibinfo {author} {\bibfnamefont {R.-Z.}\ \bibnamefont {Liu}}, \bibinfo {author} {\bibfnamefont {J.-Y.}\ \bibnamefont {Zhao}}, \bibinfo {author} {\bibfnamefont {G.-Y.}\ \bibnamefont {Zou}}, \bibinfo {author} {\bibfnamefont {H.}~\bibnamefont {Wang}}, \bibinfo {author} {\bibfnamefont {Y.}~\bibnamefont {Cao}}, \bibinfo {author} {\bibfnamefont {Y.-M.}\ \bibnamefont {He}}, \bibinfo {author} {\bibfnamefont {C.-Z.}\ \bibnamefont {Peng}}, \bibinfo {author} {\bibfnamefont {Y.-H.}\ \bibnamefont {Huo}}, \bibinfo
  {author} {\bibfnamefont {S.-K.}\ \bibnamefont {Liao}}, \bibinfo {author} {\bibfnamefont {C.-Y.}\ \bibnamefont {Lu}}, \bibinfo {author} {\bibfnamefont {F.}~\bibnamefont {Xu}}, \ and\ \bibinfo {author} {\bibfnamefont {J.-W.}\ \bibnamefont {Pan}},\ }\href {\doibase 10.1103/PhysRevLett.134.210801} {\bibfield  {journal} {\bibinfo  {journal} {Phys. Rev. Lett.}\ }\textbf {\bibinfo {volume} {134}},\ \bibinfo {pages} {210801} (\bibinfo {year} {2025})}\BibitemShut {NoStop}%
\bibitem [{\citenamefont {Bozzio}\ \emph {et~al.}(2022)\citenamefont {Bozzio}, \citenamefont {Vyvlecka}, \citenamefont {Cosacchi}, \citenamefont {Nawrath}, \citenamefont {Seidelmann}, \citenamefont {Loredo}, \citenamefont {Portalupi}, \citenamefont {Axt}, \citenamefont {Michler},\ and\ \citenamefont {Walther}}]{Bozzio2022}%
  \BibitemOpen
  \bibfield  {author} {\bibinfo {author} {\bibfnamefont {M.}~\bibnamefont {Bozzio}}, \bibinfo {author} {\bibfnamefont {M.}~\bibnamefont {Vyvlecka}}, \bibinfo {author} {\bibfnamefont {M.}~\bibnamefont {Cosacchi}}, \bibinfo {author} {\bibfnamefont {C.}~\bibnamefont {Nawrath}}, \bibinfo {author} {\bibfnamefont {T.}~\bibnamefont {Seidelmann}}, \bibinfo {author} {\bibfnamefont {J.~C.}\ \bibnamefont {Loredo}}, \bibinfo {author} {\bibfnamefont {S.~L.}\ \bibnamefont {Portalupi}}, \bibinfo {author} {\bibfnamefont {V.~M.}\ \bibnamefont {Axt}}, \bibinfo {author} {\bibfnamefont {P.}~\bibnamefont {Michler}}, \ and\ \bibinfo {author} {\bibfnamefont {P.}~\bibnamefont {Walther}},\ }\href {\doibase 10.1038/s41534-022-00626-z} {\bibfield  {journal} {\bibinfo  {journal} {npj Quantum Information}\ }\textbf {\bibinfo {volume} {8}},\ \bibinfo {pages} {104} (\bibinfo {year} {2022})}\BibitemShut {NoStop}%
\bibitem [{\citenamefont {Zou}\ \emph {et~al.}(2025)\citenamefont {Zou}, \citenamefont {He}, \citenamefont {Huang}, \citenamefont {Zhao}, \citenamefont {Li}, \citenamefont {Guo}, \citenamefont {Ding}, \citenamefont {Xu}, \citenamefont {Liu}, \citenamefont {Zou} \emph {et~al.}}]{zou2025realization}%
  \BibitemOpen
  \bibfield  {author} {\bibinfo {author} {\bibfnamefont {M.}~\bibnamefont {Zou}}, \bibinfo {author} {\bibfnamefont {Y.-M.}\ \bibnamefont {He}}, \bibinfo {author} {\bibfnamefont {Y.}~\bibnamefont {Huang}}, \bibinfo {author} {\bibfnamefont {J.-Y.}\ \bibnamefont {Zhao}}, \bibinfo {author} {\bibfnamefont {B.-C.}\ \bibnamefont {Li}}, \bibinfo {author} {\bibfnamefont {Y.-P.}\ \bibnamefont {Guo}}, \bibinfo {author} {\bibfnamefont {X.}~\bibnamefont {Ding}}, \bibinfo {author} {\bibfnamefont {M.-C.}\ \bibnamefont {Xu}}, \bibinfo {author} {\bibfnamefont {R.-Z.}\ \bibnamefont {Liu}}, \bibinfo {author} {\bibfnamefont {G.-Y.}\ \bibnamefont {Zou}},  \emph {et~al.},\ }\href@noop {} {\bibfield  {journal} {\bibinfo  {journal} {Nature Physics}\ ,\ \bibinfo {pages} {1}} (\bibinfo {year} {2025})}\BibitemShut {NoStop}%
\bibitem [{\citenamefont {Borregaard}\ \emph {et~al.}(2020)\citenamefont {Borregaard}, \citenamefont {Pichler}, \citenamefont {Schr{\"o}der}, \citenamefont {Lukin}, \citenamefont {Lodahl},\ and\ \citenamefont {S{\o}rensen}}]{borregaard2020one}%
  \BibitemOpen
  \bibfield  {author} {\bibinfo {author} {\bibfnamefont {J.}~\bibnamefont {Borregaard}}, \bibinfo {author} {\bibfnamefont {H.}~\bibnamefont {Pichler}}, \bibinfo {author} {\bibfnamefont {T.}~\bibnamefont {Schr{\"o}der}}, \bibinfo {author} {\bibfnamefont {M.~D.}\ \bibnamefont {Lukin}}, \bibinfo {author} {\bibfnamefont {P.}~\bibnamefont {Lodahl}}, \ and\ \bibinfo {author} {\bibfnamefont {A.~S.}\ \bibnamefont {S{\o}rensen}},\ }\href@noop {} {\bibfield  {journal} {\bibinfo  {journal} {Physical Review X}\ }\textbf {\bibinfo {volume} {10}},\ \bibinfo {pages} {021071} (\bibinfo {year} {2020})}\BibitemShut {NoStop}%
\bibitem [{\citenamefont {Proietti}\ \emph {et~al.}(2021)\citenamefont {Proietti}, \citenamefont {Ho}, \citenamefont {Grasselli}, \citenamefont {Barrow}, \citenamefont {Malik},\ and\ \citenamefont {Fedrizzi}}]{Proietti2021}%
  \BibitemOpen
  \bibfield  {author} {\bibinfo {author} {\bibfnamefont {M.}~\bibnamefont {Proietti}}, \bibinfo {author} {\bibfnamefont {J.}~\bibnamefont {Ho}}, \bibinfo {author} {\bibfnamefont {F.}~\bibnamefont {Grasselli}}, \bibinfo {author} {\bibfnamefont {P.}~\bibnamefont {Barrow}}, \bibinfo {author} {\bibfnamefont {M.}~\bibnamefont {Malik}}, \ and\ \bibinfo {author} {\bibfnamefont {A.}~\bibnamefont {Fedrizzi}},\ }\href {\doibase 10.1126/sciadv.abe0395} {\bibfield  {journal} {\bibinfo  {journal} {Science Advances}\ }\textbf {\bibinfo {volume} {7}},\ \bibinfo {pages} {eabe0395} (\bibinfo {year} {2021})}\BibitemShut {NoStop}%
\bibitem [{\citenamefont {Ko{\l}ody{\'n}ski}\ \emph {et~al.}(2020)\citenamefont {Ko{\l}ody{\'n}ski}, \citenamefont {M{\'a}ttar}, \citenamefont {Skrzypczyk}, \citenamefont {Woodhead}, \citenamefont {Cavalcanti}, \citenamefont {Banaszek},\ and\ \citenamefont {Ac{\'\i}n}}]{kolodynski2020device}%
  \BibitemOpen
  \bibfield  {author} {\bibinfo {author} {\bibfnamefont {J.}~\bibnamefont {Ko{\l}ody{\'n}ski}}, \bibinfo {author} {\bibfnamefont {A.}~\bibnamefont {M{\'a}ttar}}, \bibinfo {author} {\bibfnamefont {P.}~\bibnamefont {Skrzypczyk}}, \bibinfo {author} {\bibfnamefont {E.}~\bibnamefont {Woodhead}}, \bibinfo {author} {\bibfnamefont {D.}~\bibnamefont {Cavalcanti}}, \bibinfo {author} {\bibfnamefont {K.}~\bibnamefont {Banaszek}}, \ and\ \bibinfo {author} {\bibfnamefont {A.}~\bibnamefont {Ac{\'\i}n}},\ }\href@noop {} {\bibfield  {journal} {\bibinfo  {journal} {Quantum}\ }\textbf {\bibinfo {volume} {4}},\ \bibinfo {pages} {260} (\bibinfo {year} {2020})}\BibitemShut {NoStop}%
\bibitem [{\citenamefont {Gonz{\'a}lez-Ruiz}\ \emph {et~al.}(2024)\citenamefont {Gonz{\'a}lez-Ruiz}, \citenamefont {Rivera-Dean}, \citenamefont {Cenni}, \citenamefont {S{\o}rensen}, \citenamefont {Ac{\'\i}n},\ and\ \citenamefont {Oudot}}]{gonzalez2024device}%
  \BibitemOpen
  \bibfield  {author} {\bibinfo {author} {\bibfnamefont {E.~M.}\ \bibnamefont {Gonz{\'a}lez-Ruiz}}, \bibinfo {author} {\bibfnamefont {J.}~\bibnamefont {Rivera-Dean}}, \bibinfo {author} {\bibfnamefont {M.~F.}\ \bibnamefont {Cenni}}, \bibinfo {author} {\bibfnamefont {A.~S.}\ \bibnamefont {S{\o}rensen}}, \bibinfo {author} {\bibfnamefont {A.}~\bibnamefont {Ac{\'\i}n}}, \ and\ \bibinfo {author} {\bibfnamefont {E.}~\bibnamefont {Oudot}},\ }\href@noop {} {\bibfield  {journal} {\bibinfo  {journal} {Optics Express}\ }\textbf {\bibinfo {volume} {32}},\ \bibinfo {pages} {13181} (\bibinfo {year} {2024})}\BibitemShut {NoStop}%
\bibitem [{\citenamefont {Tomm}\ \emph {et~al.}(2021)\citenamefont {Tomm}, \citenamefont {Javadi}, \citenamefont {Antoniadis}, \citenamefont {Najer}, \citenamefont {L{\"o}bl}, \citenamefont {Korsch}, \citenamefont {Schott}, \citenamefont {Valentin}, \citenamefont {Wieck}, \citenamefont {Ludwig},\ and\ \citenamefont {Warburton}}]{Tomm:2021aa}%
  \BibitemOpen
  \bibfield  {author} {\bibinfo {author} {\bibfnamefont {N.}~\bibnamefont {Tomm}}, \bibinfo {author} {\bibfnamefont {A.}~\bibnamefont {Javadi}}, \bibinfo {author} {\bibfnamefont {N.~O.}\ \bibnamefont {Antoniadis}}, \bibinfo {author} {\bibfnamefont {D.}~\bibnamefont {Najer}}, \bibinfo {author} {\bibfnamefont {M.~C.}\ \bibnamefont {L{\"o}bl}}, \bibinfo {author} {\bibfnamefont {A.~R.}\ \bibnamefont {Korsch}}, \bibinfo {author} {\bibfnamefont {R.}~\bibnamefont {Schott}}, \bibinfo {author} {\bibfnamefont {S.~R.}\ \bibnamefont {Valentin}}, \bibinfo {author} {\bibfnamefont {A.~D.}\ \bibnamefont {Wieck}}, \bibinfo {author} {\bibfnamefont {A.}~\bibnamefont {Ludwig}}, \ and\ \bibinfo {author} {\bibfnamefont {R.~J.}\ \bibnamefont {Warburton}},\ }\href {\doibase 10.1038/s41565-020-00831-x} {\bibfield  {journal} {\bibinfo  {journal} {Nature Nanotechnology}\ }\textbf {\bibinfo {volume} {16}},\ \bibinfo {pages} {399} (\bibinfo {year} {2021})}\BibitemShut {NoStop}%
\bibitem [{\citenamefont {Ding}\ \emph {et~al.}(2025)\citenamefont {Ding}, \citenamefont {Guo}, \citenamefont {Xu}, \citenamefont {Liu}, \citenamefont {Zou}, \citenamefont {Zhao}, \citenamefont {Ge}, \citenamefont {Zhang}, \citenamefont {Liu}, \citenamefont {Wang}, \citenamefont {Chen}, \citenamefont {Wang}, \citenamefont {He}, \citenamefont {Huo}, \citenamefont {Lu},\ and\ \citenamefont {Pan}}]{Ding2025}%
  \BibitemOpen
  \bibfield  {author} {\bibinfo {author} {\bibfnamefont {X.}~\bibnamefont {Ding}}, \bibinfo {author} {\bibfnamefont {Y.-P.}\ \bibnamefont {Guo}}, \bibinfo {author} {\bibfnamefont {M.-C.}\ \bibnamefont {Xu}}, \bibinfo {author} {\bibfnamefont {R.-Z.}\ \bibnamefont {Liu}}, \bibinfo {author} {\bibfnamefont {G.-Y.}\ \bibnamefont {Zou}}, \bibinfo {author} {\bibfnamefont {J.-Y.}\ \bibnamefont {Zhao}}, \bibinfo {author} {\bibfnamefont {Z.-X.}\ \bibnamefont {Ge}}, \bibinfo {author} {\bibfnamefont {Q.-H.}\ \bibnamefont {Zhang}}, \bibinfo {author} {\bibfnamefont {H.-L.}\ \bibnamefont {Liu}}, \bibinfo {author} {\bibfnamefont {L.-J.}\ \bibnamefont {Wang}}, \bibinfo {author} {\bibfnamefont {M.-C.}\ \bibnamefont {Chen}}, \bibinfo {author} {\bibfnamefont {H.}~\bibnamefont {Wang}}, \bibinfo {author} {\bibfnamefont {Y.-M.}\ \bibnamefont {He}}, \bibinfo {author} {\bibfnamefont {Y.-H.}\ \bibnamefont {Huo}}, \bibinfo {author} {\bibfnamefont {C.-Y.}\ \bibnamefont {Lu}}, \ and\ \bibinfo {author} {\bibfnamefont {J.-W.}\ \bibnamefont
  {Pan}},\ }\href {\doibase 10.1038/s41566-025-01639-8} {\bibfield  {journal} {\bibinfo  {journal} {Nature Photonics}\ }\textbf {\bibinfo {volume} {19}},\ \bibinfo {pages} {387} (\bibinfo {year} {2025})}\BibitemShut {NoStop}%
\bibitem [{\citenamefont {Cao}\ \emph {et~al.}(2024)\citenamefont {Cao}, \citenamefont {Hansen}, \citenamefont {Giorgino}, \citenamefont {Carosini}, \citenamefont {Zah\'alka}, \citenamefont {Zilk}, \citenamefont {Loredo},\ and\ \citenamefont {Walther}}]{Cao2024}%
  \BibitemOpen
  \bibfield  {author} {\bibinfo {author} {\bibfnamefont {H.}~\bibnamefont {Cao}}, \bibinfo {author} {\bibfnamefont {L.~M.}\ \bibnamefont {Hansen}}, \bibinfo {author} {\bibfnamefont {F.}~\bibnamefont {Giorgino}}, \bibinfo {author} {\bibfnamefont {L.}~\bibnamefont {Carosini}}, \bibinfo {author} {\bibfnamefont {P.}~\bibnamefont {Zah\'alka}}, \bibinfo {author} {\bibfnamefont {F.}~\bibnamefont {Zilk}}, \bibinfo {author} {\bibfnamefont {J.~C.}\ \bibnamefont {Loredo}}, \ and\ \bibinfo {author} {\bibfnamefont {P.}~\bibnamefont {Walther}},\ }\href {\doibase 10.1103/PhysRevLett.132.130604} {\bibfield  {journal} {\bibinfo  {journal} {Phys. Rev. Lett.}\ }\textbf {\bibinfo {volume} {132}},\ \bibinfo {pages} {130604} (\bibinfo {year} {2024})}\BibitemShut {NoStop}%
\bibitem [{\citenamefont {Müller}\ \emph {et~al.}(2018)\citenamefont {Müller}, \citenamefont {Skiba-Szymanska}, \citenamefont {Krysa}, \citenamefont {Huwer}, \citenamefont {Felle}, \citenamefont {Anderson}, \citenamefont {Stevenson}, \citenamefont {Heffernan}, \citenamefont {Ritchie},\ and\ \citenamefont {Shields}}]{Mueller2018}%
  \BibitemOpen
  \bibfield  {author} {\bibinfo {author} {\bibfnamefont {T.}~\bibnamefont {Müller}}, \bibinfo {author} {\bibfnamefont {J.}~\bibnamefont {Skiba-Szymanska}}, \bibinfo {author} {\bibfnamefont {A.~B.}\ \bibnamefont {Krysa}}, \bibinfo {author} {\bibfnamefont {J.}~\bibnamefont {Huwer}}, \bibinfo {author} {\bibfnamefont {M.}~\bibnamefont {Felle}}, \bibinfo {author} {\bibfnamefont {M.}~\bibnamefont {Anderson}}, \bibinfo {author} {\bibfnamefont {R.~M.}\ \bibnamefont {Stevenson}}, \bibinfo {author} {\bibfnamefont {J.}~\bibnamefont {Heffernan}}, \bibinfo {author} {\bibfnamefont {D.~A.}\ \bibnamefont {Ritchie}}, \ and\ \bibinfo {author} {\bibfnamefont {A.~J.}\ \bibnamefont {Shields}},\ }\href {\doibase 10.1038/s41467-018-03251-7} {\bibfield  {journal} {\bibinfo  {journal} {Nature Communications}\ }\textbf {\bibinfo {volume} {9}},\ \bibinfo {pages} {862} (\bibinfo {year} {2018})}\BibitemShut {NoStop}%
\bibitem [{\citenamefont {Yu}\ \emph {et~al.}(2023)\citenamefont {Yu}, \citenamefont {Liu}, \citenamefont {Lee}, \citenamefont {Michler}, \citenamefont {Reitzenstein}, \citenamefont {Srinivasan}, \citenamefont {Waks},\ and\ \citenamefont {Liu}}]{Yu2023}%
  \BibitemOpen
  \bibfield  {author} {\bibinfo {author} {\bibfnamefont {Y.}~\bibnamefont {Yu}}, \bibinfo {author} {\bibfnamefont {S.}~\bibnamefont {Liu}}, \bibinfo {author} {\bibfnamefont {C.-M.}\ \bibnamefont {Lee}}, \bibinfo {author} {\bibfnamefont {P.}~\bibnamefont {Michler}}, \bibinfo {author} {\bibfnamefont {S.}~\bibnamefont {Reitzenstein}}, \bibinfo {author} {\bibfnamefont {K.}~\bibnamefont {Srinivasan}}, \bibinfo {author} {\bibfnamefont {E.}~\bibnamefont {Waks}}, \ and\ \bibinfo {author} {\bibfnamefont {J.}~\bibnamefont {Liu}},\ }\href {\doibase 10.1038/s41565-023-01528-7} {\bibfield  {journal} {\bibinfo  {journal} {Nature Nanotechnology}\ }\textbf {\bibinfo {volume} {18}},\ \bibinfo {pages} {1389} (\bibinfo {year} {2023})}\BibitemShut {NoStop}%
\bibitem [{\citenamefont {Kim}\ \emph {et~al.}(2025)\citenamefont {Kim}, \citenamefont {Kaupp}, \citenamefont {Reum}, \citenamefont {Peniakov}, \citenamefont {Michl}, \citenamefont {Kohr}, \citenamefont {Emmerling}, \citenamefont {Kamp}, \citenamefont {Cho}, \citenamefont {Huber-Loyola}, \citenamefont {Höfling},\ and\ \citenamefont {Pfenning}}]{Jewon2025}%
  \BibitemOpen
  \bibfield  {author} {\bibinfo {author} {\bibfnamefont {J.}~\bibnamefont {Kim}}, \bibinfo {author} {\bibfnamefont {J.}~\bibnamefont {Kaupp}}, \bibinfo {author} {\bibfnamefont {Y.}~\bibnamefont {Reum}}, \bibinfo {author} {\bibfnamefont {G.}~\bibnamefont {Peniakov}}, \bibinfo {author} {\bibfnamefont {J.}~\bibnamefont {Michl}}, \bibinfo {author} {\bibfnamefont {F.}~\bibnamefont {Kohr}}, \bibinfo {author} {\bibfnamefont {M.}~\bibnamefont {Emmerling}}, \bibinfo {author} {\bibfnamefont {M.}~\bibnamefont {Kamp}}, \bibinfo {author} {\bibfnamefont {Y.-H.}\ \bibnamefont {Cho}}, \bibinfo {author} {\bibfnamefont {T.}~\bibnamefont {Huber-Loyola}}, \bibinfo {author} {\bibfnamefont {S.}~\bibnamefont {Höfling}}, \ and\ \bibinfo {author} {\bibfnamefont {A.~T.}\ \bibnamefont {Pfenning}},\ }\href {\doibase https://doi.org/10.1002/qute.202500069} {\bibfield  {journal} {\bibinfo  {journal} {Advanced Quantum Technologies}\ }\textbf {\bibinfo {volume} {n/a}},\ \bibinfo {pages} {e2500069} (\bibinfo {year} {2025})}\BibitemShut
  {NoStop}%
\bibitem [{\citenamefont {Holewa}\ \emph {et~al.}(2025)\citenamefont {Holewa}, \citenamefont {Reiserer}, \citenamefont {Heindel}, \citenamefont {Sanguinetti}, \citenamefont {Huck},\ and\ \citenamefont {Semenova}}]{Holewa2025}%
  \BibitemOpen
  \bibfield  {author} {\bibinfo {author} {\bibfnamefont {P.}~\bibnamefont {Holewa}}, \bibinfo {author} {\bibfnamefont {A.}~\bibnamefont {Reiserer}}, \bibinfo {author} {\bibfnamefont {T.}~\bibnamefont {Heindel}}, \bibinfo {author} {\bibfnamefont {S.}~\bibnamefont {Sanguinetti}}, \bibinfo {author} {\bibfnamefont {A.}~\bibnamefont {Huck}}, \ and\ \bibinfo {author} {\bibfnamefont {E.}~\bibnamefont {Semenova}},\ }\href {\doibase doi:10.1515/nanoph-2024-0747} {\bibfield  {journal} {\bibinfo  {journal} {Nanophotonics}\ }\textbf {\bibinfo {volume} {14}},\ \bibinfo {pages} {1729} (\bibinfo {year} {2025})}\BibitemShut {NoStop}%
\bibitem [{\citenamefont {Morrison}\ \emph {et~al.}(2021)\citenamefont {Morrison}, \citenamefont {Rambach}, \citenamefont {Koong}, \citenamefont {Graffitti}, \citenamefont {Thorburn}, \citenamefont {Kar}, \citenamefont {Ma}, \citenamefont {Park}, \citenamefont {Song}, \citenamefont {Stoltz}, \citenamefont {Bouwmeester}, \citenamefont {Fedrizzi},\ and\ \citenamefont {Gerardot}}]{Morrison2021}%
  \BibitemOpen
  \bibfield  {author} {\bibinfo {author} {\bibfnamefont {C.~L.}\ \bibnamefont {Morrison}}, \bibinfo {author} {\bibfnamefont {M.}~\bibnamefont {Rambach}}, \bibinfo {author} {\bibfnamefont {Z.~X.}\ \bibnamefont {Koong}}, \bibinfo {author} {\bibfnamefont {F.}~\bibnamefont {Graffitti}}, \bibinfo {author} {\bibfnamefont {F.}~\bibnamefont {Thorburn}}, \bibinfo {author} {\bibfnamefont {A.~K.}\ \bibnamefont {Kar}}, \bibinfo {author} {\bibfnamefont {Y.}~\bibnamefont {Ma}}, \bibinfo {author} {\bibfnamefont {S.-I.}\ \bibnamefont {Park}}, \bibinfo {author} {\bibfnamefont {J.~D.}\ \bibnamefont {Song}}, \bibinfo {author} {\bibfnamefont {N.~G.}\ \bibnamefont {Stoltz}}, \bibinfo {author} {\bibfnamefont {D.}~\bibnamefont {Bouwmeester}}, \bibinfo {author} {\bibfnamefont {A.}~\bibnamefont {Fedrizzi}}, \ and\ \bibinfo {author} {\bibfnamefont {B.~D.}\ \bibnamefont {Gerardot}},\ }\href {\doibase 10.1063/5.0045413} {\bibfield  {journal} {\bibinfo  {journal} {Applied Physics Letters}\ }\textbf {\bibinfo {volume} {118}},\ \bibinfo
  {pages} {174003} (\bibinfo {year} {2021})}\BibitemShut {NoStop}%
\bibitem [{\citenamefont {Chiriano}\ \emph {et~al.}(2024)\citenamefont {Chiriano}, \citenamefont {Morrison}, \citenamefont {Ho}, \citenamefont {Jaeken},\ and\ \citenamefont {Fedrizzi}}]{Chiriano2025}%
  \BibitemOpen
  \bibfield  {author} {\bibinfo {author} {\bibfnamefont {F.}~\bibnamefont {Chiriano}}, \bibinfo {author} {\bibfnamefont {C.~L.}\ \bibnamefont {Morrison}}, \bibinfo {author} {\bibfnamefont {J.}~\bibnamefont {Ho}}, \bibinfo {author} {\bibfnamefont {T.}~\bibnamefont {Jaeken}}, \ and\ \bibinfo {author} {\bibfnamefont {A.}~\bibnamefont {Fedrizzi}},\ }\href {\doibase 10.1088/2058-9565/ad7f82} {\bibfield  {journal} {\bibinfo  {journal} {Quantum Science and Technology}\ }\textbf {\bibinfo {volume} {10}},\ \bibinfo {pages} {015004} (\bibinfo {year} {2024})}\BibitemShut {NoStop}%
\bibitem [{\citenamefont {Acconcia}\ \emph {et~al.}(2023)\citenamefont {Acconcia}, \citenamefont {Ceccarelli}, \citenamefont {Gulinatti},\ and\ \citenamefont {Rech}}]{Acconcia2023}%
  \BibitemOpen
  \bibfield  {author} {\bibinfo {author} {\bibfnamefont {G.}~\bibnamefont {Acconcia}}, \bibinfo {author} {\bibfnamefont {F.}~\bibnamefont {Ceccarelli}}, \bibinfo {author} {\bibfnamefont {A.}~\bibnamefont {Gulinatti}}, \ and\ \bibinfo {author} {\bibfnamefont {I.}~\bibnamefont {Rech}},\ }\href {\doibase 10.1364/OE.491400} {\bibfield  {journal} {\bibinfo  {journal} {Opt. Express}\ }\textbf {\bibinfo {volume} {31}},\ \bibinfo {pages} {33963} (\bibinfo {year} {2023})}\BibitemShut {NoStop}%
\bibitem [{\citenamefont {Ferreira~da Silva}\ \emph {et~al.}(2014)\citenamefont {Ferreira~da Silva}, \citenamefont {Xavier}, \citenamefont {Temporão},\ and\ \citenamefont {von~der Weid}}]{daSilva2014}%
  \BibitemOpen
  \bibfield  {author} {\bibinfo {author} {\bibfnamefont {T.}~\bibnamefont {Ferreira~da Silva}}, \bibinfo {author} {\bibfnamefont {G.~B.}\ \bibnamefont {Xavier}}, \bibinfo {author} {\bibfnamefont {G.~P.}\ \bibnamefont {Temporão}}, \ and\ \bibinfo {author} {\bibfnamefont {J.~P.}\ \bibnamefont {von~der Weid}},\ }\href {\doibase 10.1109/JLT.2014.2322108} {\bibfield  {journal} {\bibinfo  {journal} {Journal of Lightwave Technology}\ }\textbf {\bibinfo {volume} {32}},\ \bibinfo {pages} {2332} (\bibinfo {year} {2014})}\BibitemShut {NoStop}%
\bibitem [{\citenamefont {Mao}\ \emph {et~al.}(2018)\citenamefont {Mao}, \citenamefont {Wang}, \citenamefont {Zhao}, \citenamefont {Wang}, \citenamefont {Wang}, \citenamefont {Wang}, \citenamefont {Zhou}, \citenamefont {Nie}, \citenamefont {Chen}, \citenamefont {Zhao}, \citenamefont {Zhang}, \citenamefont {Zhang}, \citenamefont {Chen},\ and\ \citenamefont {Pan}}]{Mao2018}%
  \BibitemOpen
  \bibfield  {author} {\bibinfo {author} {\bibfnamefont {Y.}~\bibnamefont {Mao}}, \bibinfo {author} {\bibfnamefont {B.-X.}\ \bibnamefont {Wang}}, \bibinfo {author} {\bibfnamefont {C.}~\bibnamefont {Zhao}}, \bibinfo {author} {\bibfnamefont {G.}~\bibnamefont {Wang}}, \bibinfo {author} {\bibfnamefont {R.}~\bibnamefont {Wang}}, \bibinfo {author} {\bibfnamefont {H.}~\bibnamefont {Wang}}, \bibinfo {author} {\bibfnamefont {F.}~\bibnamefont {Zhou}}, \bibinfo {author} {\bibfnamefont {J.}~\bibnamefont {Nie}}, \bibinfo {author} {\bibfnamefont {Q.}~\bibnamefont {Chen}}, \bibinfo {author} {\bibfnamefont {Y.}~\bibnamefont {Zhao}}, \bibinfo {author} {\bibfnamefont {Q.}~\bibnamefont {Zhang}}, \bibinfo {author} {\bibfnamefont {J.}~\bibnamefont {Zhang}}, \bibinfo {author} {\bibfnamefont {T.-Y.}\ \bibnamefont {Chen}}, \ and\ \bibinfo {author} {\bibfnamefont {J.-W.}\ \bibnamefont {Pan}},\ }\href {\doibase 10.1364/OE.26.006010} {\bibfield  {journal} {\bibinfo  {journal} {Opt. Express}\ }\textbf {\bibinfo {volume} {26}},\ \bibinfo
  {pages} {6010} (\bibinfo {year} {2018})}\BibitemShut {NoStop}%
\bibitem [{\citenamefont {Clivati}\ \emph {et~al.}(2022)\citenamefont {Clivati}, \citenamefont {Meda}, \citenamefont {Donadello}, \citenamefont {Virzì}, \citenamefont {Genovese}, \citenamefont {Levi}, \citenamefont {Mura}, \citenamefont {Pittaluga}, \citenamefont {Yuan}, \citenamefont {Shields}, \citenamefont {Lucamarini}, \citenamefont {Degiovanni},\ and\ \citenamefont {Calonico}}]{Clivati2022}%
  \BibitemOpen
  \bibfield  {author} {\bibinfo {author} {\bibfnamefont {C.}~\bibnamefont {Clivati}}, \bibinfo {author} {\bibfnamefont {A.}~\bibnamefont {Meda}}, \bibinfo {author} {\bibfnamefont {S.}~\bibnamefont {Donadello}}, \bibinfo {author} {\bibfnamefont {S.}~\bibnamefont {Virzì}}, \bibinfo {author} {\bibfnamefont {M.}~\bibnamefont {Genovese}}, \bibinfo {author} {\bibfnamefont {F.}~\bibnamefont {Levi}}, \bibinfo {author} {\bibfnamefont {A.}~\bibnamefont {Mura}}, \bibinfo {author} {\bibfnamefont {M.}~\bibnamefont {Pittaluga}}, \bibinfo {author} {\bibfnamefont {Z.}~\bibnamefont {Yuan}}, \bibinfo {author} {\bibfnamefont {A.~J.}\ \bibnamefont {Shields}}, \bibinfo {author} {\bibfnamefont {M.}~\bibnamefont {Lucamarini}}, \bibinfo {author} {\bibfnamefont {I.~P.}\ \bibnamefont {Degiovanni}}, \ and\ \bibinfo {author} {\bibfnamefont {D.}~\bibnamefont {Calonico}},\ }\href {\doibase 10.1038/s41467-021-27808-1} {\bibfield  {journal} {\bibinfo  {journal} {Nature Communications}\ }\textbf {\bibinfo {volume} {13}},\ \bibinfo {pages}
  {157} (\bibinfo {year} {2022})}\BibitemShut {NoStop}%
\bibitem [{\citenamefont {Grünenfelder}\ \emph {et~al.}(2021)\citenamefont {Grünenfelder}, \citenamefont {Sax}, \citenamefont {Boaron},\ and\ \citenamefont {Zbinden}}]{Grünenfelder2021}%
  \BibitemOpen
  \bibfield  {author} {\bibinfo {author} {\bibfnamefont {F.}~\bibnamefont {Grünenfelder}}, \bibinfo {author} {\bibfnamefont {R.}~\bibnamefont {Sax}}, \bibinfo {author} {\bibfnamefont {A.}~\bibnamefont {Boaron}}, \ and\ \bibinfo {author} {\bibfnamefont {H.}~\bibnamefont {Zbinden}},\ }\href {\doibase 10.1063/5.0060232} {\bibfield  {journal} {\bibinfo  {journal} {Applied Physics Letters}\ }\textbf {\bibinfo {volume} {119}},\ \bibinfo {pages} {124001} (\bibinfo {year} {2021})}\BibitemShut {NoStop}%
\bibitem [{\citenamefont {Petrovich}\ \emph {et~al.}(2025)\citenamefont {Petrovich}, \citenamefont {Numkam~Fokoua}, \citenamefont {Chen}, \citenamefont {Sakr}, \citenamefont {Adamu}, \citenamefont {Hassan}, \citenamefont {Wu}, \citenamefont {Fatobene~Ando}, \citenamefont {Papadimopoulos}, \citenamefont {Sandoghchi}, \citenamefont {Jasion},\ and\ \citenamefont {Poletti}}]{Petrovich2025}%
  \BibitemOpen
  \bibfield  {author} {\bibinfo {author} {\bibfnamefont {M.}~\bibnamefont {Petrovich}}, \bibinfo {author} {\bibfnamefont {E.}~\bibnamefont {Numkam~Fokoua}}, \bibinfo {author} {\bibfnamefont {Y.}~\bibnamefont {Chen}}, \bibinfo {author} {\bibfnamefont {H.}~\bibnamefont {Sakr}}, \bibinfo {author} {\bibfnamefont {A.~I.}\ \bibnamefont {Adamu}}, \bibinfo {author} {\bibfnamefont {R.}~\bibnamefont {Hassan}}, \bibinfo {author} {\bibfnamefont {D.}~\bibnamefont {Wu}}, \bibinfo {author} {\bibfnamefont {R.}~\bibnamefont {Fatobene~Ando}}, \bibinfo {author} {\bibfnamefont {A.}~\bibnamefont {Papadimopoulos}}, \bibinfo {author} {\bibfnamefont {S.~R.}\ \bibnamefont {Sandoghchi}}, \bibinfo {author} {\bibfnamefont {G.}~\bibnamefont {Jasion}}, \ and\ \bibinfo {author} {\bibfnamefont {F.}~\bibnamefont {Poletti}},\ }\href {\doibase 10.1038/s41566-025-01747-5} {\bibfield  {journal} {\bibinfo  {journal} {Nature Photonics}\ } (\bibinfo {year} {2025}),\ 10.1038/s41566-025-01747-5}\BibitemShut {NoStop}%
\bibitem [{\citenamefont {Alia}\ \emph {et~al.}(2022)\citenamefont {Alia}, \citenamefont {Tessinari}, \citenamefont {Bahrani}, \citenamefont {Bradley}, \citenamefont {Sakr}, \citenamefont {Harrington}, \citenamefont {Hayes}, \citenamefont {Chen}, \citenamefont {Petropoulos}, \citenamefont {Richardson}, \citenamefont {Poletti}, \citenamefont {Kanellos}, \citenamefont {Nejabati},\ and\ \citenamefont {Simeonidou}}]{Alia2022}%
  \BibitemOpen
  \bibfield  {author} {\bibinfo {author} {\bibfnamefont {O.}~\bibnamefont {Alia}}, \bibinfo {author} {\bibfnamefont {R.~S.}\ \bibnamefont {Tessinari}}, \bibinfo {author} {\bibfnamefont {S.}~\bibnamefont {Bahrani}}, \bibinfo {author} {\bibfnamefont {T.~D.}\ \bibnamefont {Bradley}}, \bibinfo {author} {\bibfnamefont {H.}~\bibnamefont {Sakr}}, \bibinfo {author} {\bibfnamefont {K.}~\bibnamefont {Harrington}}, \bibinfo {author} {\bibfnamefont {J.}~\bibnamefont {Hayes}}, \bibinfo {author} {\bibfnamefont {Y.}~\bibnamefont {Chen}}, \bibinfo {author} {\bibfnamefont {P.}~\bibnamefont {Petropoulos}}, \bibinfo {author} {\bibfnamefont {D.}~\bibnamefont {Richardson}}, \bibinfo {author} {\bibfnamefont {F.}~\bibnamefont {Poletti}}, \bibinfo {author} {\bibfnamefont {G.~T.}\ \bibnamefont {Kanellos}}, \bibinfo {author} {\bibfnamefont {R.}~\bibnamefont {Nejabati}}, \ and\ \bibinfo {author} {\bibfnamefont {D.}~\bibnamefont {Simeonidou}},\ }\href {\doibase 10.1109/JLT.2022.3180232} {\bibfield  {journal} {\bibinfo  {journal}
  {Journal of Lightwave Technology}\ }\textbf {\bibinfo {volume} {40}},\ \bibinfo {pages} {5522} (\bibinfo {year} {2022})}\BibitemShut {NoStop}%
\bibitem [{\citenamefont {Honz}\ \emph {et~al.}(2022)\citenamefont {Honz}, \citenamefont {Prawits}, \citenamefont {Alia}, \citenamefont {Sakr}, \citenamefont {Bradley}, \citenamefont {Zhang}, \citenamefont {Slavik}, \citenamefont {Poletti}, \citenamefont {Kanellos}, \citenamefont {Nejabati}, \citenamefont {Walther}, \citenamefont {Simeonidou}, \citenamefont {H{\"u}bel},\ and\ \citenamefont {Schrenk}}]{Honz2022}%
  \BibitemOpen
  \bibfield  {author} {\bibinfo {author} {\bibfnamefont {F.}~\bibnamefont {Honz}}, \bibinfo {author} {\bibfnamefont {F.}~\bibnamefont {Prawits}}, \bibinfo {author} {\bibfnamefont {O.}~\bibnamefont {Alia}}, \bibinfo {author} {\bibfnamefont {H.}~\bibnamefont {Sakr}}, \bibinfo {author} {\bibfnamefont {T.}~\bibnamefont {Bradley}}, \bibinfo {author} {\bibfnamefont {C.}~\bibnamefont {Zhang}}, \bibinfo {author} {\bibfnamefont {R.}~\bibnamefont {Slavik}}, \bibinfo {author} {\bibfnamefont {F.}~\bibnamefont {Poletti}}, \bibinfo {author} {\bibfnamefont {G.}~\bibnamefont {Kanellos}}, \bibinfo {author} {\bibfnamefont {R.}~\bibnamefont {Nejabati}}, \bibinfo {author} {\bibfnamefont {P.}~\bibnamefont {Walther}}, \bibinfo {author} {\bibfnamefont {D.}~\bibnamefont {Simeonidou}}, \bibinfo {author} {\bibfnamefont {H.}~\bibnamefont {H{\"u}bel}}, \ and\ \bibinfo {author} {\bibfnamefont {B.}~\bibnamefont {Schrenk}},\ }in\ \href@noop {} {{\selectlanguage {English}\emph {\bibinfo {booktitle} {2022 European Conference on Optical
  Communication (ECOC)}}}}\ (\bibinfo {year} {2022})\ pp.\ \bibinfo {pages} {1--4},\ \bibinfo {note} {european Conference on Optical Communication (ECOC) 2022 ; Conference date: 18-09-2022 Through 22-09-2022}\BibitemShut {NoStop}%
\bibitem [{\citenamefont {Alia}\ \emph {et~al.}(2025)\citenamefont {Alia}, \citenamefont {Clark}, \citenamefont {Bahrani}, \citenamefont {Wang}, \citenamefont {Jasion}, \citenamefont {Sakr}, \citenamefont {Hayes}, \citenamefont {Petropoulos}, \citenamefont {Poletti}, \citenamefont {Kanellos}, \citenamefont {Rarity}, \citenamefont {Nejabati}, \citenamefont {Joshi},\ and\ \citenamefont {Simeonidou}}]{Alia2025}%
  \BibitemOpen
  \bibfield  {author} {\bibinfo {author} {\bibfnamefont {O.}~\bibnamefont {Alia}}, \bibinfo {author} {\bibfnamefont {M.~J.}\ \bibnamefont {Clark}}, \bibinfo {author} {\bibfnamefont {S.}~\bibnamefont {Bahrani}}, \bibinfo {author} {\bibfnamefont {R.}~\bibnamefont {Wang}}, \bibinfo {author} {\bibfnamefont {G.~T.}\ \bibnamefont {Jasion}}, \bibinfo {author} {\bibfnamefont {H.}~\bibnamefont {Sakr}}, \bibinfo {author} {\bibfnamefont {J.~R.}\ \bibnamefont {Hayes}}, \bibinfo {author} {\bibfnamefont {P.}~\bibnamefont {Petropoulos}}, \bibinfo {author} {\bibfnamefont {F.}~\bibnamefont {Poletti}}, \bibinfo {author} {\bibfnamefont {G.~T.}\ \bibnamefont {Kanellos}}, \bibinfo {author} {\bibfnamefont {J.~G.}\ \bibnamefont {Rarity}}, \bibinfo {author} {\bibfnamefont {R.}~\bibnamefont {Nejabati}}, \bibinfo {author} {\bibfnamefont {S.~K.}\ \bibnamefont {Joshi}}, \ and\ \bibinfo {author} {\bibfnamefont {D.}~\bibnamefont {Simeonidou}},\ }in\ \href {\doibase 10.1364/OFC.2025.Tu3D.1} {\emph {\bibinfo {booktitle} {Optical Fiber
  Communication Conference (OFC) 2025}}}\ (\bibinfo  {publisher} {Optica Publishing Group},\ \bibinfo {year} {2025})\ p.\ \bibinfo {pages} {Tu3D.1}\BibitemShut {NoStop}%
\bibitem [{\citenamefont {Antesberger}\ \emph {et~al.}(2024)\citenamefont {Antesberger}, \citenamefont {Richter}, \citenamefont {Poletti}, \citenamefont {Slav\'{i}k}, \citenamefont {Petropoulos}, \citenamefont {H\"{u}bel}, \citenamefont {Trenti}, \citenamefont {Walther},\ and\ \citenamefont {Rozema}}]{Antesberger2024}%
  \BibitemOpen
  \bibfield  {author} {\bibinfo {author} {\bibfnamefont {M.}~\bibnamefont {Antesberger}}, \bibinfo {author} {\bibfnamefont {C.~M.~D.}\ \bibnamefont {Richter}}, \bibinfo {author} {\bibfnamefont {F.}~\bibnamefont {Poletti}}, \bibinfo {author} {\bibfnamefont {R.}~\bibnamefont {Slav\'{i}k}}, \bibinfo {author} {\bibfnamefont {P.}~\bibnamefont {Petropoulos}}, \bibinfo {author} {\bibfnamefont {H.}~\bibnamefont {H\"{u}bel}}, \bibinfo {author} {\bibfnamefont {A.}~\bibnamefont {Trenti}}, \bibinfo {author} {\bibfnamefont {P.}~\bibnamefont {Walther}}, \ and\ \bibinfo {author} {\bibfnamefont {L.~A.}\ \bibnamefont {Rozema}},\ }\href {\doibase 10.1364/OPTICAQ.514257} {\bibfield  {journal} {\bibinfo  {journal} {Optica Quantum}\ }\textbf {\bibinfo {volume} {2}},\ \bibinfo {pages} {173} (\bibinfo {year} {2024})}\BibitemShut {NoStop}%
\bibitem [{\citenamefont {Trenti}\ \emph {et~al.}(2024)\citenamefont {Trenti}, \citenamefont {Luchian}, \citenamefont {Poletti}, \citenamefont {Slavík}, \citenamefont {Petropoulos}, \citenamefont {Alia}, \citenamefont {Kanellos},\ and\ \citenamefont {Hübel}}]{Trenti2024}%
  \BibitemOpen
  \bibfield  {author} {\bibinfo {author} {\bibfnamefont {A.}~\bibnamefont {Trenti}}, \bibinfo {author} {\bibfnamefont {C.}~\bibnamefont {Luchian}}, \bibinfo {author} {\bibfnamefont {F.}~\bibnamefont {Poletti}}, \bibinfo {author} {\bibfnamefont {R.}~\bibnamefont {Slavík}}, \bibinfo {author} {\bibfnamefont {P.}~\bibnamefont {Petropoulos}}, \bibinfo {author} {\bibfnamefont {O.}~\bibnamefont {Alia}}, \bibinfo {author} {\bibfnamefont {G.~T.}\ \bibnamefont {Kanellos}}, \ and\ \bibinfo {author} {\bibfnamefont {H.}~\bibnamefont {Hübel}},\ }\href {\doibase 10.1109/JSTQE.2024.3392416} {\bibfield  {journal} {\bibinfo  {journal} {IEEE Journal of Selected Topics in Quantum Electronics}\ }\textbf {\bibinfo {volume} {30}},\ \bibinfo {pages} {1} (\bibinfo {year} {2024})}\BibitemShut {NoStop}%
\bibitem [{\citenamefont {Minder}\ \emph {et~al.}(2023)\citenamefont {Minder}, \citenamefont {Albosh}, \citenamefont {Alia}, \citenamefont {Slavik}, \citenamefont {Kumar}, \citenamefont {Poletti}, \citenamefont {Kanellos},\ and\ \citenamefont {Lucamarini}}]{Minder2023}%
  \BibitemOpen
  \bibfield  {author} {\bibinfo {author} {\bibfnamefont {M.}~\bibnamefont {Minder}}, \bibinfo {author} {\bibfnamefont {S.}~\bibnamefont {Albosh}}, \bibinfo {author} {\bibfnamefont {O.}~\bibnamefont {Alia}}, \bibinfo {author} {\bibfnamefont {R.}~\bibnamefont {Slavik}}, \bibinfo {author} {\bibfnamefont {R.}~\bibnamefont {Kumar}}, \bibinfo {author} {\bibfnamefont {F.}~\bibnamefont {Poletti}}, \bibinfo {author} {\bibfnamefont {G.}~\bibnamefont {Kanellos}}, \ and\ \bibinfo {author} {\bibfnamefont {M.}~\bibnamefont {Lucamarini}},\ }in\ \href {\doibase 10.1117/12.2647583} {\emph {\bibinfo {booktitle} {Quantum Technology: Driving Commercialisation of an Enabling Science III}}},\ Vol.\ \bibinfo {volume} {12335},\ \bibinfo {editor} {edited by\ \bibinfo {editor} {\bibfnamefont {M.~J.}\ \bibnamefont {Padgett}}, \bibinfo {editor} {\bibfnamefont {K.}~\bibnamefont {Bongs}}, \bibinfo {editor} {\bibfnamefont {A.}~\bibnamefont {Fedrizzi}}, \ and\ \bibinfo {editor} {\bibfnamefont {A.}~\bibnamefont {Politi}}},\ \bibinfo
  {organization} {International Society for Optics and Photonics}\ (\bibinfo  {publisher} {SPIE},\ \bibinfo {year} {2023})\ p.\ \bibinfo {pages} {123350I}\BibitemShut {NoStop}%
\bibitem [{\citenamefont {Poletti}(2014)}]{Poletti2014}%
  \BibitemOpen
  \bibfield  {author} {\bibinfo {author} {\bibfnamefont {F.}~\bibnamefont {Poletti}},\ }\href {\doibase 10.1364/OE.22.023807} {\bibfield  {journal} {\bibinfo  {journal} {Opt. Express}\ }\textbf {\bibinfo {volume} {22}},\ \bibinfo {pages} {23807} (\bibinfo {year} {2014})}\BibitemShut {NoStop}%
\bibitem [{\citenamefont {Suslov}\ \emph {et~al.}(2022)\citenamefont {Suslov}, \citenamefont {Fokoua}, \citenamefont {Dousek}, \citenamefont {Zhong}, \citenamefont {Zv\'{a}novec}, \citenamefont {Bradley}, \citenamefont {Poletti}, \citenamefont {Richardson}, \citenamefont {Komanec},\ and\ \citenamefont {Slav\'{i}k}}]{Suslov2022}%
  \BibitemOpen
  \bibfield  {author} {\bibinfo {author} {\bibfnamefont {D.}~\bibnamefont {Suslov}}, \bibinfo {author} {\bibfnamefont {E.~N.}\ \bibnamefont {Fokoua}}, \bibinfo {author} {\bibfnamefont {D.}~\bibnamefont {Dousek}}, \bibinfo {author} {\bibfnamefont {A.}~\bibnamefont {Zhong}}, \bibinfo {author} {\bibfnamefont {S.}~\bibnamefont {Zv\'{a}novec}}, \bibinfo {author} {\bibfnamefont {T.~D.}\ \bibnamefont {Bradley}}, \bibinfo {author} {\bibfnamefont {F.}~\bibnamefont {Poletti}}, \bibinfo {author} {\bibfnamefont {D.~J.}\ \bibnamefont {Richardson}}, \bibinfo {author} {\bibfnamefont {M.}~\bibnamefont {Komanec}}, \ and\ \bibinfo {author} {\bibfnamefont {R.}~\bibnamefont {Slav\'{i}k}},\ }\href {\doibase 10.1364/OE.460635} {\bibfield  {journal} {\bibinfo  {journal} {Opt. Express}\ }\textbf {\bibinfo {volume} {30}},\ \bibinfo {pages} {37006} (\bibinfo {year} {2022})}\BibitemShut {NoStop}%
\bibitem [{\citenamefont {Zhong}\ \emph {et~al.}(2024)\citenamefont {Zhong}, \citenamefont {Fokoua}, \citenamefont {Ding}, \citenamefont {Dousek}, \citenamefont {Suslov}, \citenamefont {Zv\'{a}novec}, \citenamefont {Poletti}, \citenamefont {Slav\'{i}k},\ and\ \citenamefont {Komanec}}]{Zhong2024}%
  \BibitemOpen
  \bibfield  {author} {\bibinfo {author} {\bibfnamefont {A.}~\bibnamefont {Zhong}}, \bibinfo {author} {\bibfnamefont {E.~N.}\ \bibnamefont {Fokoua}}, \bibinfo {author} {\bibfnamefont {M.}~\bibnamefont {Ding}}, \bibinfo {author} {\bibfnamefont {D.}~\bibnamefont {Dousek}}, \bibinfo {author} {\bibfnamefont {D.}~\bibnamefont {Suslov}}, \bibinfo {author} {\bibfnamefont {S.}~\bibnamefont {Zv\'{a}novec}}, \bibinfo {author} {\bibfnamefont {F.}~\bibnamefont {Poletti}}, \bibinfo {author} {\bibfnamefont {R.}~\bibnamefont {Slav\'{i}k}}, \ and\ \bibinfo {author} {\bibfnamefont {M.}~\bibnamefont {Komanec}},\ }\href {\doibase 10.1364/JLT.42.002124} {\bibfield  {journal} {\bibinfo  {journal} {J. Lightwave Technol.}\ }\textbf {\bibinfo {volume} {42}},\ \bibinfo {pages} {2124} (\bibinfo {year} {2024})}\BibitemShut {NoStop}%
\bibitem [{\citenamefont {Uppu}\ \emph {et~al.}(2020)\citenamefont {Uppu}, \citenamefont {Pedersen}, \citenamefont {Wang}, \citenamefont {Olesen}, \citenamefont {Papon}, \citenamefont {Zhou}, \citenamefont {Midolo}, \citenamefont {Scholz}, \citenamefont {Wieck}, \citenamefont {Ludwig},\ and\ \citenamefont {Lodahl}}]{UPPU:20}%
  \BibitemOpen
  \bibfield  {author} {\bibinfo {author} {\bibfnamefont {R.}~\bibnamefont {Uppu}}, \bibinfo {author} {\bibfnamefont {F.~T.}\ \bibnamefont {Pedersen}}, \bibinfo {author} {\bibfnamefont {Y.}~\bibnamefont {Wang}}, \bibinfo {author} {\bibfnamefont {C.~T.}\ \bibnamefont {Olesen}}, \bibinfo {author} {\bibfnamefont {C.}~\bibnamefont {Papon}}, \bibinfo {author} {\bibfnamefont {X.}~\bibnamefont {Zhou}}, \bibinfo {author} {\bibfnamefont {L.}~\bibnamefont {Midolo}}, \bibinfo {author} {\bibfnamefont {S.}~\bibnamefont {Scholz}}, \bibinfo {author} {\bibfnamefont {A.~D.}\ \bibnamefont {Wieck}}, \bibinfo {author} {\bibfnamefont {A.}~\bibnamefont {Ludwig}}, \ and\ \bibinfo {author} {\bibfnamefont {P.}~\bibnamefont {Lodahl}},\ }\href {\doibase 10.1126/sciadv.abc8268} {\bibfield  {journal} {\bibinfo  {journal} {Science Advances}\ }\textbf {\bibinfo {volume} {6}},\ \bibinfo {pages} {eabc8268} (\bibinfo {year} {2020})}\BibitemShut {NoStop}%
\bibitem [{\citenamefont {Pedersen}\ \emph {et~al.}(2020)\citenamefont {Pedersen}, \citenamefont {Wang}, \citenamefont {Olesen}, \citenamefont {Scholz}, \citenamefont {Wieck}, \citenamefont {Ludwig}, \citenamefont {LObl}, \citenamefont {Warburton}, \citenamefont {Midolo}, \citenamefont {Uppu} \emph {et~al.}}]{pedersen2020near}%
  \BibitemOpen
  \bibfield  {author} {\bibinfo {author} {\bibfnamefont {F.~T.}\ \bibnamefont {Pedersen}}, \bibinfo {author} {\bibfnamefont {Y.}~\bibnamefont {Wang}}, \bibinfo {author} {\bibfnamefont {C.~T.}\ \bibnamefont {Olesen}}, \bibinfo {author} {\bibfnamefont {S.}~\bibnamefont {Scholz}}, \bibinfo {author} {\bibfnamefont {A.~D.}\ \bibnamefont {Wieck}}, \bibinfo {author} {\bibfnamefont {A.}~\bibnamefont {Ludwig}}, \bibinfo {author} {\bibfnamefont {M.~C.}\ \bibnamefont {LObl}}, \bibinfo {author} {\bibfnamefont {R.~J.}\ \bibnamefont {Warburton}}, \bibinfo {author} {\bibfnamefont {L.}~\bibnamefont {Midolo}}, \bibinfo {author} {\bibfnamefont {R.}~\bibnamefont {Uppu}},  \emph {et~al.},\ }\href@noop {} {\bibfield  {journal} {\bibinfo  {journal} {ACS Photonics}\ }\textbf {\bibinfo {volume} {7}},\ \bibinfo {pages} {2343} (\bibinfo {year} {2020})}\BibitemShut {NoStop}%
\bibitem [{\citenamefont {Senellart}\ \emph {et~al.}(2017)\citenamefont {Senellart}, \citenamefont {Solomon},\ and\ \citenamefont {White}}]{sps:senellart17}%
  \BibitemOpen
  \bibfield  {author} {\bibinfo {author} {\bibfnamefont {P.}~\bibnamefont {Senellart}}, \bibinfo {author} {\bibfnamefont {G.}~\bibnamefont {Solomon}}, \ and\ \bibinfo {author} {\bibfnamefont {A.}~\bibnamefont {White}},\ }\href {http://dx.doi.org/10.1038/nnano.2017.218} {\bibfield  {journal} {\bibinfo  {journal} {Nature Nanotechnology}\ }\textbf {\bibinfo {volume} {12}},\ \bibinfo {pages} {1026 EP } (\bibinfo {year} {2017})}\BibitemShut {NoStop}%
\bibitem [{\citenamefont {Brown}\ and\ \citenamefont {Twiss}(1956)}]{Brown1956}%
  \BibitemOpen
  \bibfield  {author} {\bibinfo {author} {\bibfnamefont {R.~H.}\ \bibnamefont {Brown}}\ and\ \bibinfo {author} {\bibfnamefont {R.~Q.}\ \bibnamefont {Twiss}},\ }\href {\doibase 10.1038/177027a0} {\bibfield  {journal} {\bibinfo  {journal} {Nature}\ }\textbf {\bibinfo {volume} {177}},\ \bibinfo {pages} {27} (\bibinfo {year} {1956})}\BibitemShut {NoStop}%
\bibitem [{\citenamefont {Hong}\ \emph {et~al.}(1987)\citenamefont {Hong}, \citenamefont {Ou},\ and\ \citenamefont {Mandel}}]{HOM87}%
  \BibitemOpen
  \bibfield  {author} {\bibinfo {author} {\bibfnamefont {C.~K.}\ \bibnamefont {Hong}}, \bibinfo {author} {\bibfnamefont {Z.~Y.}\ \bibnamefont {Ou}}, \ and\ \bibinfo {author} {\bibfnamefont {L.}~\bibnamefont {Mandel}},\ }\href@noop {} {\bibfield  {journal} {\bibinfo  {journal} {Phys. Rev. Lett.}\ }\textbf {\bibinfo {volume} {59}},\ \bibinfo {pages} {2044} (\bibinfo {year} {1987})}\BibitemShut {NoStop}%
\bibitem [{\citenamefont {Loredo}\ \emph {et~al.}(2016)\citenamefont {Loredo}, \citenamefont {Zakaria}, \citenamefont {Somaschi}, \citenamefont {Anton}, \citenamefont {de~Santis}, \citenamefont {Giesz}, \citenamefont {Grange}, \citenamefont {Broome}, \citenamefont {Gazzano}, \citenamefont {Coppola}, \citenamefont {Sagnes}, \citenamefont {Lemaitre}, \citenamefont {Auffeves}, \citenamefont {Senellart}, \citenamefont {Almeida},\ and\ \citenamefont {White}}]{Loredo:16}%
  \BibitemOpen
  \bibfield  {author} {\bibinfo {author} {\bibfnamefont {J.~C.}\ \bibnamefont {Loredo}}, \bibinfo {author} {\bibfnamefont {N.~A.}\ \bibnamefont {Zakaria}}, \bibinfo {author} {\bibfnamefont {N.}~\bibnamefont {Somaschi}}, \bibinfo {author} {\bibfnamefont {C.}~\bibnamefont {Anton}}, \bibinfo {author} {\bibfnamefont {L.}~\bibnamefont {de~Santis}}, \bibinfo {author} {\bibfnamefont {V.}~\bibnamefont {Giesz}}, \bibinfo {author} {\bibfnamefont {T.}~\bibnamefont {Grange}}, \bibinfo {author} {\bibfnamefont {M.~A.}\ \bibnamefont {Broome}}, \bibinfo {author} {\bibfnamefont {O.}~\bibnamefont {Gazzano}}, \bibinfo {author} {\bibfnamefont {G.}~\bibnamefont {Coppola}}, \bibinfo {author} {\bibfnamefont {I.}~\bibnamefont {Sagnes}}, \bibinfo {author} {\bibfnamefont {A.}~\bibnamefont {Lemaitre}}, \bibinfo {author} {\bibfnamefont {A.}~\bibnamefont {Auffeves}}, \bibinfo {author} {\bibfnamefont {P.}~\bibnamefont {Senellart}}, \bibinfo {author} {\bibfnamefont {M.~P.}\ \bibnamefont {Almeida}}, \ and\ \bibinfo {author} {\bibfnamefont
  {A.~G.}\ \bibnamefont {White}},\ }\href {\doibase 10.1364/OPTICA.3.000433} {\bibfield  {journal} {\bibinfo  {journal} {Optica}\ }\textbf {\bibinfo {volume} {3}},\ \bibinfo {pages} {433} (\bibinfo {year} {2016})}\BibitemShut {NoStop}%
\bibitem [{\citenamefont {Ollivier}\ \emph {et~al.}(2021)\citenamefont {Ollivier}, \citenamefont {Thomas}, \citenamefont {Wein}, \citenamefont {de~Buy~Wenniger}, \citenamefont {Coste}, \citenamefont {Loredo}, \citenamefont {Somaschi}, \citenamefont {Harouri}, \citenamefont {Lemaitre}, \citenamefont {Sagnes}, \citenamefont {Lanco}, \citenamefont {Simon}, \citenamefont {Anton}, \citenamefont {Krebs},\ and\ \citenamefont {Senellart}}]{Ollivier2021}%
  \BibitemOpen
  \bibfield  {author} {\bibinfo {author} {\bibfnamefont {H.}~\bibnamefont {Ollivier}}, \bibinfo {author} {\bibfnamefont {S.}~\bibnamefont {Thomas}}, \bibinfo {author} {\bibfnamefont {S.}~\bibnamefont {Wein}}, \bibinfo {author} {\bibfnamefont {I.~M.}\ \bibnamefont {de~Buy~Wenniger}}, \bibinfo {author} {\bibfnamefont {N.}~\bibnamefont {Coste}}, \bibinfo {author} {\bibfnamefont {J.}~\bibnamefont {Loredo}}, \bibinfo {author} {\bibfnamefont {N.}~\bibnamefont {Somaschi}}, \bibinfo {author} {\bibfnamefont {A.}~\bibnamefont {Harouri}}, \bibinfo {author} {\bibfnamefont {A.}~\bibnamefont {Lemaitre}}, \bibinfo {author} {\bibfnamefont {I.}~\bibnamefont {Sagnes}}, \bibinfo {author} {\bibfnamefont {L.}~\bibnamefont {Lanco}}, \bibinfo {author} {\bibfnamefont {C.}~\bibnamefont {Simon}}, \bibinfo {author} {\bibfnamefont {C.}~\bibnamefont {Anton}}, \bibinfo {author} {\bibfnamefont {O.}~\bibnamefont {Krebs}}, \ and\ \bibinfo {author} {\bibfnamefont {P.}~\bibnamefont {Senellart}},\ }\href {\doibase
  10.1103/physrevlett.126.063602} {\bibfield  {journal} {\bibinfo  {journal} {Physical Review Letters}\ }\textbf {\bibinfo {volume} {126}} (\bibinfo {year} {2021}),\ 10.1103/physrevlett.126.063602}\BibitemShut {NoStop}%
\bibitem [{\citenamefont {González-Ruiz}\ \emph {et~al.}(2025)\citenamefont {González-Ruiz}, \citenamefont {Bjerlin}, \citenamefont {Sandberg},\ and\ \citenamefont {Sørensen}}]{GonzlezRuiz2025}%
  \BibitemOpen
  \bibfield  {author} {\bibinfo {author} {\bibfnamefont {E.~M.}\ \bibnamefont {González-Ruiz}}, \bibinfo {author} {\bibfnamefont {J.}~\bibnamefont {Bjerlin}}, \bibinfo {author} {\bibfnamefont {O.~A.~D.}\ \bibnamefont {Sandberg}}, \ and\ \bibinfo {author} {\bibfnamefont {A.~S.}\ \bibnamefont {Sørensen}},\ }\href {\doibase 10.1103/physrevapplied.23.054063} {\bibfield  {journal} {\bibinfo  {journal} {Physical Review Applied}\ }\textbf {\bibinfo {volume} {23}} (\bibinfo {year} {2025}),\ 10.1103/physrevapplied.23.054063}\BibitemShut {NoStop}%
\bibitem [{\citenamefont {Gottesman}\ \emph {et~al.}(2004)\citenamefont {Gottesman}, \citenamefont {Lo}, \citenamefont {Lutkenhaus},\ and\ \citenamefont {Preskill}}]{Gottesman2004}%
  \BibitemOpen
  \bibfield  {author} {\bibinfo {author} {\bibfnamefont {D.}~\bibnamefont {Gottesman}}, \bibinfo {author} {\bibfnamefont {H.-K.}\ \bibnamefont {Lo}}, \bibinfo {author} {\bibfnamefont {N.}~\bibnamefont {Lutkenhaus}}, \ and\ \bibinfo {author} {\bibfnamefont {J.}~\bibnamefont {Preskill}},\ }in\ \href {\doibase 10.1109/isit.2004.1365172} {\emph {\bibinfo {booktitle} {International Symposium onInformation Theory, 2004. ISIT 2004. Proceedings.}}}\ (\bibinfo  {publisher} {IEEE},\ \bibinfo {year} {2004})\ p.\ \bibinfo {pages} {135–135}\BibitemShut {NoStop}%
\bibitem [{\citenamefont {Morrison}\ \emph {et~al.}(2023)\citenamefont {Morrison}, \citenamefont {Pousa}, \citenamefont {Graffitti}, \citenamefont {Koong}, \citenamefont {Barrow}, \citenamefont {Stoltz}, \citenamefont {Bouwmeester}, \citenamefont {Jeffers}, \citenamefont {Oi}, \citenamefont {Gerardot},\ and\ \citenamefont {Fedrizzi}}]{Morrison2023}%
  \BibitemOpen
  \bibfield  {author} {\bibinfo {author} {\bibfnamefont {C.~L.}\ \bibnamefont {Morrison}}, \bibinfo {author} {\bibfnamefont {R.~G.}\ \bibnamefont {Pousa}}, \bibinfo {author} {\bibfnamefont {F.}~\bibnamefont {Graffitti}}, \bibinfo {author} {\bibfnamefont {Z.~X.}\ \bibnamefont {Koong}}, \bibinfo {author} {\bibfnamefont {P.}~\bibnamefont {Barrow}}, \bibinfo {author} {\bibfnamefont {N.~G.}\ \bibnamefont {Stoltz}}, \bibinfo {author} {\bibfnamefont {D.}~\bibnamefont {Bouwmeester}}, \bibinfo {author} {\bibfnamefont {J.}~\bibnamefont {Jeffers}}, \bibinfo {author} {\bibfnamefont {D.~K.~L.}\ \bibnamefont {Oi}}, \bibinfo {author} {\bibfnamefont {B.~D.}\ \bibnamefont {Gerardot}}, \ and\ \bibinfo {author} {\bibfnamefont {A.}~\bibnamefont {Fedrizzi}},\ }\href {\doibase 10.1038/s41467-023-39219-5} {\bibfield  {journal} {\bibinfo  {journal} {Nature Communications}\ }\textbf {\bibinfo {volume} {14}} (\bibinfo {year} {2023}),\ 10.1038/s41467-023-39219-5}\BibitemShut {NoStop}%
\bibitem [{\citenamefont {Zahidy}\ \emph {et~al.}(2024)\citenamefont {Zahidy}, \citenamefont {Mikkelsen}, \citenamefont {M\"{u}ller}, \citenamefont {Da~Lio}, \citenamefont {Krehbiel}, \citenamefont {Wang}, \citenamefont {Bart}, \citenamefont {Wieck}, \citenamefont {Ludwig}, \citenamefont {Galili}, \citenamefont {Forchhammer}, \citenamefont {Lodahl}, \citenamefont {Oxenløwe}, \citenamefont {Bacco},\ and\ \citenamefont {Midolo}}]{Zahidy2024}%
  \BibitemOpen
  \bibfield  {author} {\bibinfo {author} {\bibfnamefont {M.}~\bibnamefont {Zahidy}}, \bibinfo {author} {\bibfnamefont {M.~T.}\ \bibnamefont {Mikkelsen}}, \bibinfo {author} {\bibfnamefont {R.}~\bibnamefont {M\"{u}ller}}, \bibinfo {author} {\bibfnamefont {B.}~\bibnamefont {Da~Lio}}, \bibinfo {author} {\bibfnamefont {M.}~\bibnamefont {Krehbiel}}, \bibinfo {author} {\bibfnamefont {Y.}~\bibnamefont {Wang}}, \bibinfo {author} {\bibfnamefont {N.}~\bibnamefont {Bart}}, \bibinfo {author} {\bibfnamefont {A.~D.}\ \bibnamefont {Wieck}}, \bibinfo {author} {\bibfnamefont {A.}~\bibnamefont {Ludwig}}, \bibinfo {author} {\bibfnamefont {M.}~\bibnamefont {Galili}}, \bibinfo {author} {\bibfnamefont {S.}~\bibnamefont {Forchhammer}}, \bibinfo {author} {\bibfnamefont {P.}~\bibnamefont {Lodahl}}, \bibinfo {author} {\bibfnamefont {L.~K.}\ \bibnamefont {Oxenløwe}}, \bibinfo {author} {\bibfnamefont {D.}~\bibnamefont {Bacco}}, \ and\ \bibinfo {author} {\bibfnamefont {L.}~\bibnamefont {Midolo}},\ }\href {\doibase
  10.1038/s41534-023-00800-x} {\bibfield  {journal} {\bibinfo  {journal} {npj Quantum Information}\ }\textbf {\bibinfo {volume} {10}} (\bibinfo {year} {2024}),\ 10.1038/s41534-023-00800-x}\BibitemShut {NoStop}%
\bibitem [{\citenamefont {Fokoua}\ \emph {et~al.}(2023)\citenamefont {Fokoua}, \citenamefont {Mousavi}, \citenamefont {Jasion}, \citenamefont {Richardson},\ and\ \citenamefont {Poletti}}]{Fokoua2023}%
  \BibitemOpen
  \bibfield  {author} {\bibinfo {author} {\bibfnamefont {E.~N.}\ \bibnamefont {Fokoua}}, \bibinfo {author} {\bibfnamefont {S.~A.}\ \bibnamefont {Mousavi}}, \bibinfo {author} {\bibfnamefont {G.~T.}\ \bibnamefont {Jasion}}, \bibinfo {author} {\bibfnamefont {D.~J.}\ \bibnamefont {Richardson}}, \ and\ \bibinfo {author} {\bibfnamefont {F.}~\bibnamefont {Poletti}},\ }\href {\doibase 10.1364/AOP.470592} {\bibfield  {journal} {\bibinfo  {journal} {Adv. Opt. Photon.}\ }\textbf {\bibinfo {volume} {15}},\ \bibinfo {pages} {1} (\bibinfo {year} {2023})}\BibitemShut {NoStop}%
\end{thebibliography}%

\end{document}